\documentclass{appolb}

\usepackage{amssymb}
\usepackage{amsmath}
\usepackage{epsfig}

\def \et {E_{T}}
\newcommand{\met}{\mbox{$\not\!\!\et$}}

\newcommand{\specialcell}[2][c]{%
  \begin{tabular}[#1]{@{}c@{}}#2\end{tabular}}
\DeclareRobustCommand{\fbi}{\ensuremath{\mathrm{fb}^{-1}}}

\newcommand{\be}{\begin{equation}}
\newcommand{\ee}{\end{equation}}
\newcommand{\bea}{\begin{eqnarray}}
\newcommand{\eea}{\end{eqnarray}}

\def\gev{\, {\rm GeV}}

\begin{document}


\title{Forward physics at the LHC: from the structure of the Pomeron to the
search for $\gamma$-induced resonances}

\author{Christophe Royon 
\address{The University of Kansas, Lawrence, USA \\
email: christophe.royon@ku.edu} }


\maketitle

\begin{abstract}
We describe some of the future measurements to be performed by the CMS, TOTEM
and ATLAS
collaborations on hard diffraction in order to understand better the
structure of the Pomeron. We also describe the prospects concerning the
search for quartic $\gamma \gamma \gamma \gamma$ anomalous couplings and discuss
a possible
interpretation for the existence of a new particle decaying into two photons at a mass
of about 750 GeV.                                                   
\end{abstract}

In this short review, we will describe some potential measurements to be performed
at the LHC mainly in the ATLAS, CMS-TOTEM, CT-PPS experiments in order to get a better
understanding of diffraction and photon-exchange processes. Of special interest
will be the discussion of beyond standard model reaches especially in the
di-photon channel that might be the firt sign of new physics. These studies follow
a long term collaboration with Prof. Andrzej Bialas and Prof. Robert Peschanski 
that started after my PhD in Saclay about the dipole model and
diffraction~\cite{dipole} and I would like to express all my gratitude to Andrezj
for this long term and successful collaboration, and to wish him a very nice
birthday for this occasion.

\section{Experimental definition of diffraction and measurement of the gluon density in the
Pomeron at HERA}

In this section, we discuss the different experimental ways to define
diffraction. As an example, we describe the methods used by the H1 and ZEUS
experiments at HERA, DESY, Hamburg in Germany since it is the starting point for
any diffraction studies at the LHC (most of the quark and gluon densities in the
Pomeron were obtained using HERA data). In addition, many results concerning
diffraction have been obtained at the Tevatron and the LHC will allow to extend
these measurements in a completely new kinematical domain.

\subsection{The rapidity gap method}
\begin{figure}[t]
\begin{center}
\vspace{10.cm}
\hspace{-4cm}
\epsfig{file=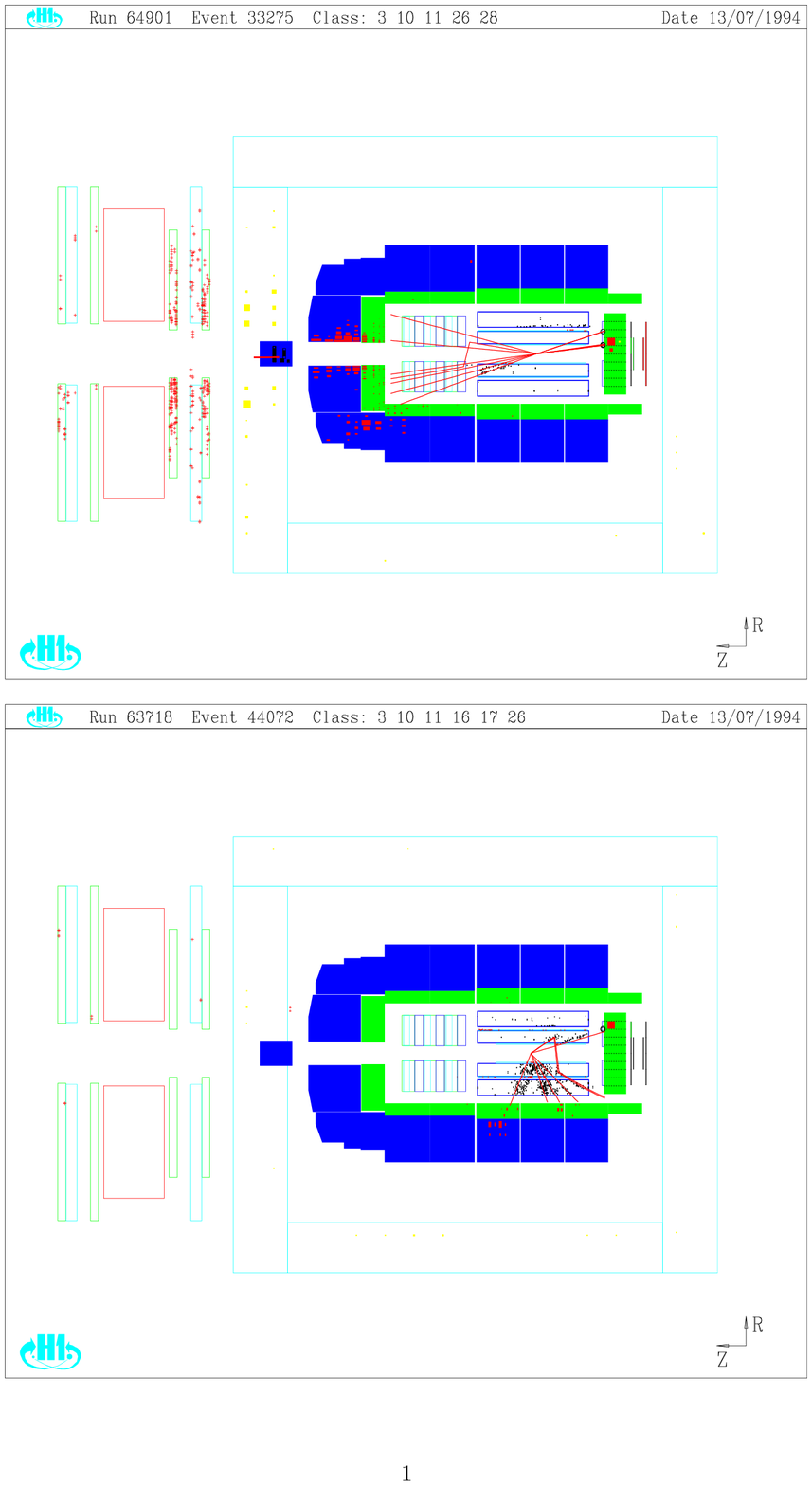,width=5.2cm}
\caption{``Usual" and diffractive events in the H1 experiment.\label{fig1}
}
\end{center}
\end{figure}

HERA is a collider where electrons of 27.6 GeV collide with protons of 920 GeV.
A typical event as shown in the upper plot of Fig. 1 is $ep \rightarrow eX$
where electron and jets are produced in the final state. We
notice that the electron is scattered in the H1 backward detector\footnote{At
HERA, the backward (resp. forward) directions are defined as the direction
of the outgoing electron (resp. proton).} (in green)
whereas some hadronic activity is present in the forward region of the detector
(in the LAr calorimeter and in the forward muon detectors). The proton is thus
completely destroyed and the interaction leads to jets and proton remnants directly observable
in the detector. The fact that much energy is observed in the forward region is
due to colour exchange between the scattered jet and the proton remnants.
In about 10\% of the events, the situation is completely
different. Such events appear like the one shown in the bottom plot of
Fig.~\ref{fig1}.
The electron is still present in the backward detector, there is
still some hadronic activity (jets) in the LAr calorimeter, but no energy above
noise level is deposited in the forward part of the LAr calorimeter or in the
forward muon detectors. In other words, there is no color exchange between the
proton and the produced jets. As an example, this can be explained if the proton stays intact
after the interaction. 

This experimental observation leads to the first definition of diffraction:
request a
rapidity gap (in other words a domain in the forward detectors where  no
energy is deposited above noise level) in the forward region. For example, the H1
collaboration requests no energy deposition in the rapidity region
$3.3 < \eta < 7.5$ where $\eta$ is the pseudorapidity. Let us note that this
approach does not insure that the proton stays intact after the interaction, but
it represents a limit on the mass of the produced object $M_Y<1.6$ GeV. Within
this limit, the proton could be dissociated. The advantage of the rapidity gap
method is that it is quite easy to implement and it has a large acceptance in 
the diffractive kinematical
plane. The inconvenient is that it is difficult to use at the LHC because of pile up 
events. In order to accumulate high luminosities at the LHC, many proton interactions
occur within the same bunch crossing and a diffractive event will be overlapping with
non-diffractive events that will induce the presence of energy in the forward region. 

\subsection{Proton tagging}
The second experimental method to detect diffractive events is also natural: the
idea is to detect directly the intact proton in the final state. The proton
loses a small fraction of its energy and is thus scattered at very small angle
with respect to the beam direction. Some special detectors called roman pots can
be used to detect the protons close to the beam. The basic idea is simple: the roman pot
detectors are located far away from the interaction point and can move close to
the beam, when the beam is stable, to detect protons scattered at vary small
angles. The inconvenience is that the kinematical reach of those detectors is
usually smaller than with the rapidity gap method. On the other hand,
the advantage is that it
gives a clear signal of diffraction since it measures the diffracted proton
directly.

\begin{figure}[t]
\begin{center}
\epsfig{file=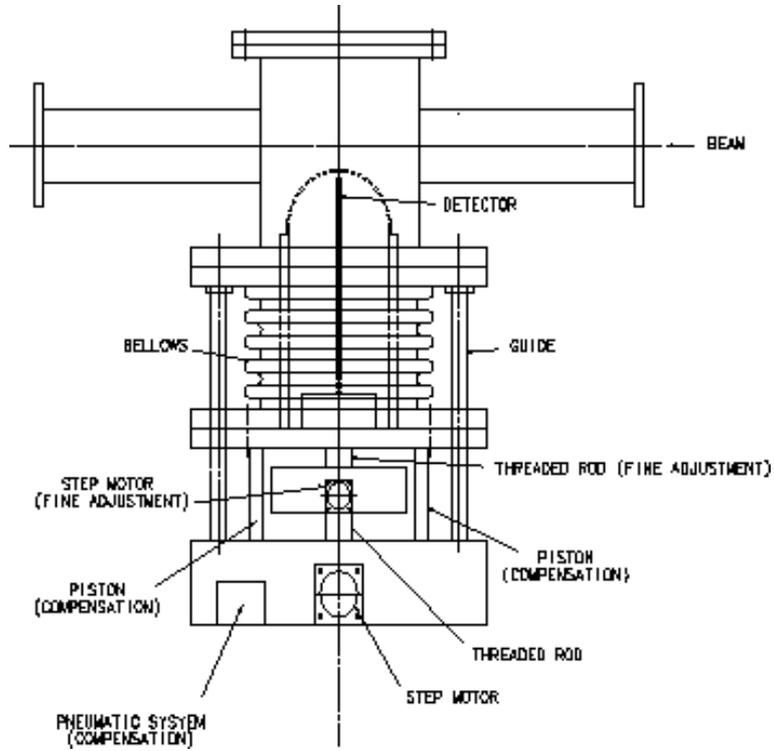,width=10cm}
\caption{Scheme of a roman pot detector.\label{fig2}
}
\end{center}
\end{figure}

A scheme of a roman pot detector as it is used by the H1 or ZEUS experiment is shown
in Fig.~\ref{fig2}, and similar detectors are used by the TOTEM, CMS-TOTEM and 
ATLAS collaborations at the LHC. The beam is the horizontal line at the upper part of the
figure. The detector is located in the pot itself and can move closer to the
beam when the beam is stable enough (during the injection period, the detectors
are protected in the home position). Step motors allow to move the detectors
with high precision. A precise knowledge  of the detector position is
necessary to reconstruct the transverse momentum of the scattered proton and
thus the diffractive kinematical variables. The detectors are placed in a
secondary vaccuum with respect to the beam one. 

\subsection{Diffractive kinematical variables}
\begin{figure}[t]
\begin{center}
\epsfig{file=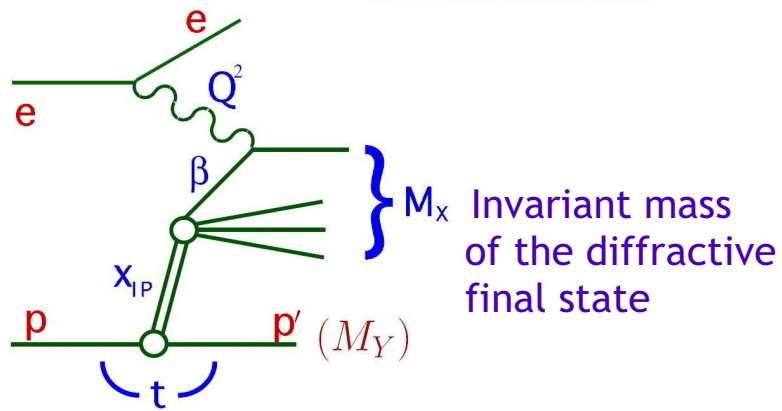,width=6cm,angle=270}
\caption{Scheme of a diffractive event at HERA.\label{fig4}
}
\end{center}
\end{figure}

After having described the different experimental definitions of diffraction at
HERA, we will give the new kinematical variables used to characterise diffraction.
A typical diffractive event is shown in Fig.~\ref{fig4} where $ep \rightarrow
epX$ is depicted. In addition to the usual deep inelastic variables, $Q^2$ the transfered
energy squared at the electron vertex, $x$ the fraction of the proton momentum
carried by the struck quark, 
$W^2 = Q^2 (1/x -1)$ the total energy in the final state,
new diffractive variables are defined: $x_P$ (called $\xi$ at the Tevatron
and the LHC) is the
momentum fraction of the proton carried by the colourless object called the
pomeron, and $\beta$ the momentum fraction of the pomeron carried by the
interacting parton inside the pomeron if we assume the pomeron to be made of
quarks and gluons.
\begin{eqnarray}
x_P &=& \xi = \frac{Q^2+M_X^2}{Q^2+W^2} \\
\beta &=& \frac{Q^2}{Q^2+M_X^2} = \frac{x}{x_P}.
\end{eqnarray}

\subsection{Extraction of the gluon density in the Pomeron}

The general idea is to measure the cross section to produce diffractive events 
either by requesting the presence of a rapidity gap in the forward direction or
of a tagged proton in roman pot detectors as a function of the $t$, $\xi$, $\beta$, $Q^2$
kinematic variables.
The following step is to perform Dokshitzer Gribov Lipatov Altarelli Parisi (DGLAP)
\cite{dglap} fits to the pomeron structure function. If we assume that the
pomeron is made of quarks and gluons, it is natural to check whether the DGLAP
evolution equations are able to describe the $Q^2$ evolution of these parton
densities.

\begin{figure}[t]
\begin{center}
\epsfig{file=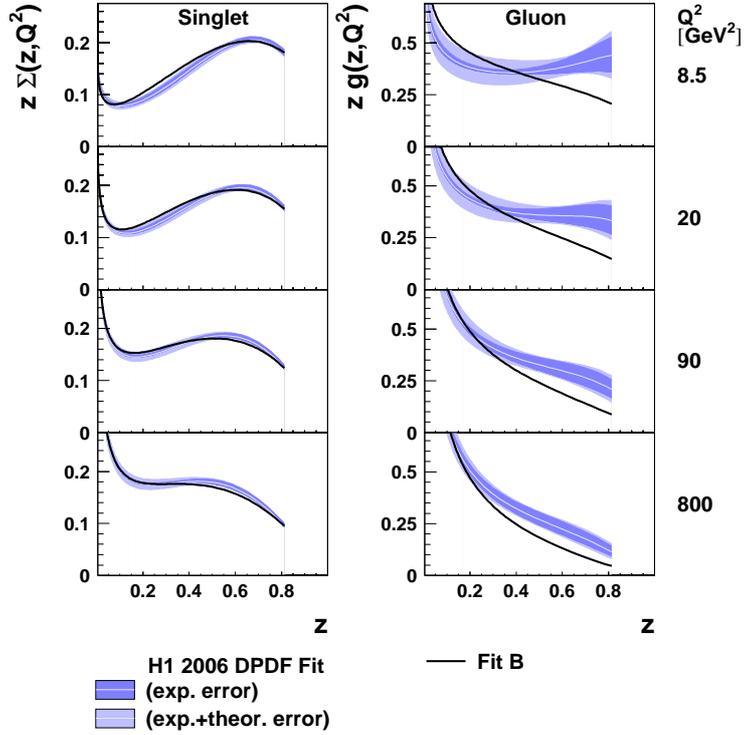,width=10cm}
\vspace{12cm}
\caption{Extraction of the parton densities in the pomeron using a DGLAP NLO fit
(H1 collaboration).\label{fig8}
}
\end{center}
\end{figure}

The DGLAP QCD fit allows to get the parton distributions in the pomeron as a
direct output of the fit~\cite{fith1}, and they are displayed in Fig.~\ref{fig8} as a blue shaded
area as a function of $\beta$. We first note that the gluon density is much
higher than the quark one, showing that the pomeron is gluon dominated. We also
note that the gluon density at high $\beta$ is poorly constrained which is shown
by the larger shaded area. The measurement of dijet cross section in diffraction
allows contraining the gluon density further but the high $\beta$ density is
still poorly constrained.

\section{Diffraction at the LHC}

In the same way that we discuss diffraction at HERA, diffraction can occur at 
the LHC, the 13 TeV $p p$ collider located close to Geneva, at CERN,
Switzerland.  In that case, one can have diffraction on one side only (single
diffraction) or on both sides (double pomeron exchange). 

\subsection{Diffractive kinematical variables}

As we just mentioned, diffraction at the LHC can occur on both $p$ sides.
In the same way
as we defined the kinematical variables $x_P$ and $\beta$ at HERA, we define $\xi_{1,2}$(=$x_P$ at HERA) 
as the proton fractional momentum loss (or as the $p$ 
momentum fraction carried by the pomeron), and $\beta_{1,2}$, the fraction of the
pomeron momentum carried by the interacting parton. The produced diffractive
mass is equal to $M^2= s \xi_1 $ for single diffractive events and to
$M^2= s \xi_1 \xi_2$ for double pomeron exchange. The size of the rapidity gap
is of the order of $\Delta \eta \sim \log 1/ \xi_{1,2}$.

The rapidity gap method can be only used at low luminosity at the LHC. At high
instantaneous luminosity, many interactions (called pile up) occur within the
same bunch crossing. The pile up interactions will fill in the rapidity gap 
devoid of any energy, making difficult to use the rapidity gap method. It is
thus preferable to tag directly the protons at the LHC.

\subsection{Diffraction at the LHC}
In this short report we discuss some potential measurements that can be
accomplished in forward physics at the LHC. We distinguish between the low
luminosity (no pile up), medium luminosity (moderate pile up) and high
luminosity (high pile up) environments. Forward physics is fundamental at the
LHC since it adresses the QCD dynamics at the interface between hard and soft
physics. For instance, the soft total $pp$ cross section probes long transverse
distances, and the BFKL~\cite{bfkl} pomeron is valid at short distances. 
In addition,
diffraction and especially photon exchange processes allow performing searches
beyond the standard model. Diffractive events are also important to
tune MC and understand underlying events and soft QCD. 
More details about the different measurements
can be found in~\cite{yellow}.

\subsection{LHC running conditions and forward detectors}

\subsubsection{Forward detectors}
At the LHC, the different detectors are sensitive to different programs of
forward physics. The LHCf detector~\cite{lhcf} measures the multiplicities and energy flow
in the very forward direction at very low
luminosity. The selection of diffractive events in LHCb~\cite{lhcb} and 
Alice~\cite{alice} is performed
by using the so-called rapidity gap method and will benefit from new 
scintillators
that cover the forward region as was installed previously in CMS. The present
coverage of the CMS and ATLAS forward detectors is complemented by the AFP
and CMS-TOTEM/CT-PPS projects to add additional proton detectors at about 220
meters from the interaction point~\cite{nicolo, projects}. 

\begin{figure}
\begin{center}
\epsfig{file=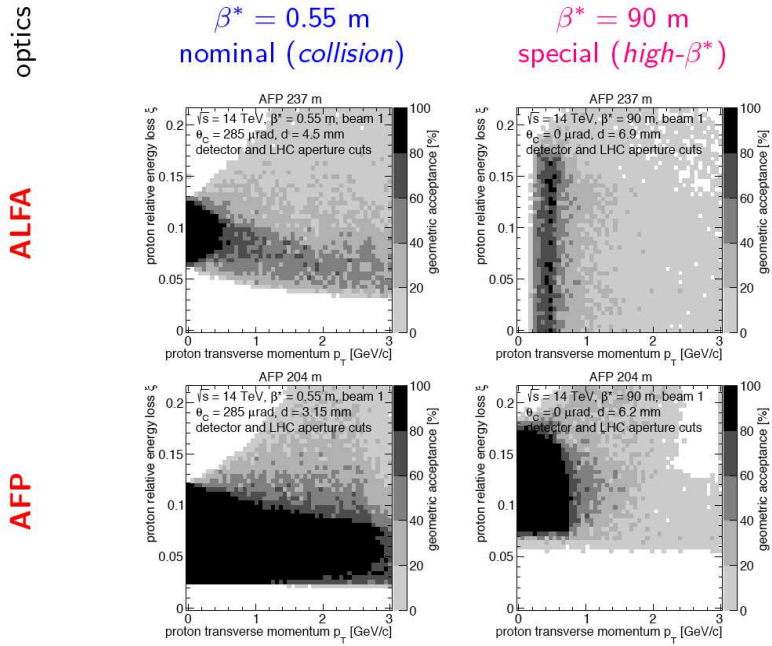,width=10.5cm}
\caption{Acceptance $\xi$ versus $t$ at low and high $\beta^*$ for vertical
(ALFA) and horizontal (AFP) roman pots at 220 m.}
\label{kinatlas}
\end{center}
\end{figure}

Running at low and high $\beta^*$ using the CMS-TOTEM, CT-PPS and ATLAS-AFP
detectors allows accessing different kinematical domains for diffraction.
In Fig.~\ref{kinatlas} are displayed the acceptances in proton relative energy
loss $\xi$ versus the proton transverse momentum $p_T$ for two values of
$\beta^*$ (0.55 m, the nominal collision optics, and 90 m) for vertical (ALFA)
or horizontal (AFP) roman pot detector configurations located about 220 m from
the ATLAS interaction point~\cite{yellow}. We notice that one can access low and high mass
diffraction (low and high $\xi$) at high $\beta^*$ in ALFA and only low mass
diffraction (up to $\xi \sim$0.15) at low $\beta^*$ using AFP. Both measurements
will be thus interesting in order to cover easily low and high mass
diffraction. The kinematical coverage is similar for the vertical (CMS-TOTEM)
and the horizontal pots (CT-PPS) of CMS and TOTEM.

\subsubsection{Different luminosity conditions}

As we mentioned in the last section, we distinguish between the low, medium
and high luminosity runs~\cite{reviewdif}.

The low luminosity runs (without pile up) allow performing multiplicity and
energy flow measurements useful to tune MC as well as to measure the total and
soft diffractive cross sections in the ATLAS/ALFA and TOTEM
experiments. Additional measurements such as single diffraction, low mass
resonances and glueballs typically require a few days of data taking (0.1 to 1.
pb$^{-1}$).

Medium luminosity runs are specific for the different LHC experiments. LHCb
accumulate typically a few fb$^{-1}$ at low pile up during their nominal data
taking while the CMS-TOTEM and ATLAS (ALFA and AFP) can accumulate low pile up
data in low and high $\beta^*$ special runs at low luminosity at the LHC. 
It is then typically possible to accumulate
1 to 10 pb$^{-1}$ at high $\beta^*$ with a pile up $\mu \sim$1 with a few days
of data taking and 10 to 100 pb$^{-1}$ at low $\beta^*$ with one to two weeks of
data taking at $\mu \sim$2 to 5.

High pile up data taking means taking all the luminosity delivered typically to
ATLAS and CMS with a pile up $\mu$ between 20 and 100. It is also possible to
collect data at a lower pile up $\mu \sim$25 by restricting to end of store data
taking (up to 40\% of the total luminosity can be collected in this way).
or to data originating from the tails of the vertex distribution.
 
\subsection{Low luminosity measurements}

In addition to measurements of the total and soft diffraction cross sections
performed at high $\beta^*$ in dedicated runs, 
data taken without pile up are specially interesting to measure multiplicities
and energy flow especially useful to tune MC benefitting from the different coverage
in rapidity of the different LHC experiments. There is also a special interest
driven by the cosmic ray community to measure the multiplicities in
proton-oxygen runs at the LHC since models make different predictions in those
conditions even if they lead to similar predictions in proton proton interactions
at 14 TeV. This will allow making precise predictions on proton oxygen events
for cosmic ray physics.

\begin{figure}
\begin{center}
\epsfig{file=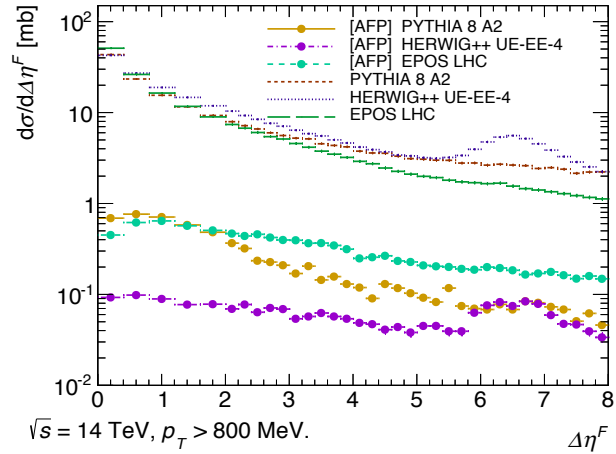,width=8.5cm}
\caption{Size of rapidity in diffractive events  for different MC models 
when protons are tagged in AFP or not.}
\label{tuning}
\end{center}
\end{figure}

Another example of fundamental measurements to be performed at very low
luminosity is the measurement of the size of the forward gap in diffractive
events when the protons are tagged in AFP or in TOTEM. The differences between
the models are much larger when the protons are tagged~\cite{yellow}, and this will allow
further tuning of the models as shown in Fig.~\ref{tuning}.

\subsection{Medium luminosity measurements}

\subsubsection{Inclusive diffractive measurements}
Medium luminosity measurements with the rapidity gap method used in Alice (two
new scintillator hodoscopes covering $-7.0< \eta < -4.9$ and $4.8 < \eta < 6.3$
are being installed in Alice in order to improve the forward coverage) or with
proton tagging in AFP and CMS-TOTEM allow constraining further the pomeron
structure using $\gamma+$jet and dijet events~\cite{qcd}. The aim is to
answer mainly the following questions that are fundamental from the QCD point of
view:
\begin{itemize}
\item Is it the same object (the same pomeron) which explains diffraction in
$pp$ (LHC) and $ep$ (HERA)? Are the measurements compatible between the
different accelerators?
\item If yes, what are the further constraints of the pomeron structure in terms
of quarks and gluons?
\item What is the value of the survival probability? It is important to measure
it since it is difficult to compute it theoretically, being sensitive to
non-perturbative physics
\end{itemize}

Feasibility studies have been performed in ATLAS (and measurements started in
CMS-TOTEM at 8 TeV) concerning the possibility to measure jet production cross
sections in single diffractive and double pomeron exchange events at low
$\beta^*$~\cite{yellow}.

\subsubsection{Dijet production in double Pomeron exchanges processes and sensitivity to the gluon
density in the pomeron}

\begin{figure}
\begin{center}
\epsfig{file=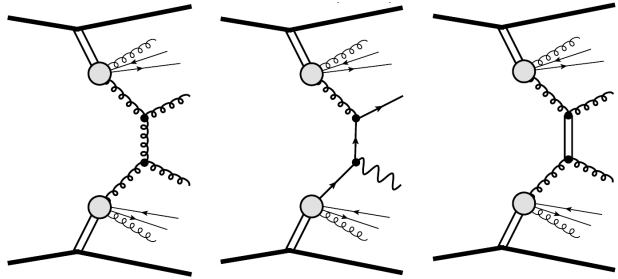,width=8.5cm}
\caption{Inclusive diffractive diagrams. From left to right: jet production in inclusive double
pomeron exchange, $\gamma +$jet production in DPE, jet gap jet events}
\label{d0b}
\end{center}
\end{figure}

One can first probe if the Pomeron is universal between
$ep$ and $pp$ colliders, or in 
other words, if we are sensitive to the same object at HERA and the LHC using as an
example dijet production in single
diffractive and double pomeron exchange at the LHC. 
It is possible to assess the gluon and quark densities 
using the dijet and $\gamma + jet$ productions.
The different diagrams of the processes that can be studied at the LHC
are shown in Fig.~\ref{d0b}, namely double pomeron exchange (DPE) production of dijets (left),
of $\gamma +$jet (middle), sensitive respectively to the gluon and quark contents of the
Pomeron, and the jet gap jet events (right). 

\begin{figure}
\begin{center}
\epsfig{file=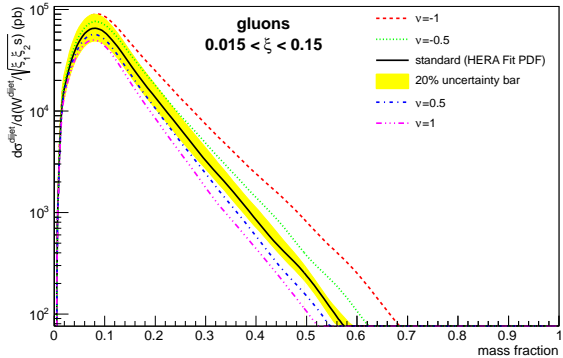,width=8.3cm}
\caption{DPE di-jet mass fraction distribution. The different curves correspond to different 
modifications of the Pomeron gluon density extracted from HERA data (see text).}
\label{fig6b}
\end{center}
\end{figure}

The dijet production in DPE events at the LHC is sensitive to the gluon density
in the Pomeron. In order to
quantify how well we are sensitive to the Pomeron structure in terms of gluon
density at the LHC, we display in Fig.~\ref{fig6b}, the dijet mass fraction,
the ratio of the dijet mass to the total diffractive mass~\cite{qcd,matthias,cdfus}.
The central black line displays the cross
section value for the gluon density in the Pomeron measured at HERA including an
additional survival probability of 0.03. The yellow band shows the effect of the
20\% uncertainty on the gluon density taking into account the normalisation
uncertainties. The dashed curves display how the dijet
cross section at the LHC is sensitive to the gluon density distribution
especially at high $\beta$. For this sake, we multiply the gluon density in the
Pomeron from HERA by $(1-\beta)^{\nu}$ where $\nu$ varies between -1 and 1. When
$\nu$ is equal to -1 (resp. 1), the gluon density is enhanced (resp, decreased)
at high $\beta$. We note
that the curves corresponding to the different values of $\nu$ are much more 
separated at high values of the dijet mass fraction, meaning that this observable
is indeed sensitive to the gluon density at high $\beta$.

\subsubsection{Sensitivity to the Pomeron structure in quarks using 
$\gamma + \textnormal{jet}$ events and $W$ asymmetry}

The QCD diffractive fits performed at HERA assumed that
$u=d=s=\bar{u}=\bar{d}=\bar{s}$, since data were not sensitive to the
difference between the different quark component in the Pomeron. 
On the contrary, measuring the $\gamma +$jet to the
dijet cross section ratios as a function of the diffractive mass $M$ allows to
distinguish between different
assumptions on the quark content of the Pomeron~\cite{qcd}. For instance, varying $d/u$ between 0.25 and
4  leads to a variation of the cross section ratio by a factor
2.5. These measuremnts can be performed both at low and high $\beta^*$ leading to
different kinematical domains in jet and photon $p_T$.

In addition, it is possible to use the $W$ asymmetry in single diffractive $W$
production~\cite{annabelle}. Typically, the muon asymmetry is directly sensitive to the
quark content of the pomeron and varies by a factor 6 at low $\xi$ between the
assumptions $u/d=2$ or $u/d=1/2$ for the quark content in the pomeron.

\subsection{Probing BFKL dynamics in diffractive events}

In this subsection, we will discuss how one can probe NLL BFKL resummation 
effects using gap between jets events at the LHC in double pomeron exchanges
(we will assume that the proton can be tagged in AFP or CMS/TOTEM).

The production cross section of two jets with a gap in rapidity between them 
reads
\begin{equation}
\frac{d \sigma^{pp\to XJJY}}{dx_1 dx_2 dE_T^2} = {\cal S}f_{eff}(x_1,E_T^2)f_{eff}(x_2,E_T^2)
\frac{d \sigma^{gg\rightarrow gg}}{dE_T^2},
\label{jgj}\end{equation}
where $\sqrt{s}$ is the total energy of the collision,
$E_T$ the transverse momentum of the two jets, $x_1$ and $x_2$ their longitudinal
fraction of momentum with respect to the incident hadrons, $S$ the survival probability,
and $f$ the effective parton density functions~\cite{usb}. The rapidity gap
between the two jets is $\Delta\eta\!=\!\ln(x_1x_2s/p_T^2).$ 

The cross section is given by
\begin{equation}
\frac{d \sigma^{gg\rightarrow gg}}{dE_T^2}=\frac{1}{16\pi}\left|A(\Delta\eta,E_T^2)\right|^2
\end{equation}
in terms of the $gg\to gg$ scattering amplitude $A(\Delta\eta,p_T^2).$ 

In the following, we consider the high energy limit in which the rapidity gap $\Delta\eta$ is assumed to be very large.
The BFKL framework allows to compute the $gg\to gg$ amplitude in this regime, and the result is 
known up to NLL accuracy
\begin{eqnarray}
A(\Delta\eta,E_T^2) &=& \frac{16N_c\pi\alpha_s^2}{C_FE_T^2}\sum_{p=-\infty}^\infty 
\int \frac{d \gamma}{2 i \pi} A_p \\
A_p &=& \frac{[p^2-(\gamma-1/2)^2]\exp\left\{\bar\alpha(E_T^2)\chi_{eff}[2p,\gamma,\bar\alpha(E_T^2)] \Delta \eta\right\}}
{[(\gamma-1/2)^2-(p-1/2)^2][(\gamma-1/2)^2-(p+1/2)^2]} 
\label{jgjnll}
\end{eqnarray}
with the complex integral running along the imaginary axis from $1/2\!-\!i\infty$ 
to $1/2\!+\!i\infty,$ and with only even conformal spins contributing to the sum, and 
$\bar{\alpha}=\alpha_S N_C/\pi$ the running coupling.

In this study, we performed a parametrised distribution of $d \sigma^{gg\rightarrow gg}/dE_T^2$
so that it can be easily implemented in the Herwig Monte Carlo~\cite{herwig} since performing the integral over
$\gamma$ in particular would be too much time consuming in a Monte Carlo. The implementation of the
BFKL cross section in a Monte Carlo is absolutely necessary to make a direct comparison with data.
Namely, the measurements are sensititive to the jet size (for instance, experimentally the gap size
is different from the rapidity interval between the jets which is not the case by definition in the
analytic calculation).

\begin{figure}
\begin{center}
\epsfig{file=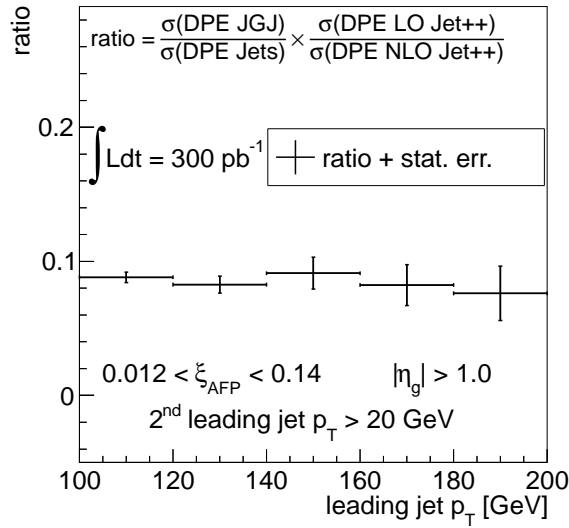,width=7.5cm}

\caption{Ratio of the jet-gap-jet to the inclusive jet cross sections at the LHC 
as a function of jet $p_T$ in double pomeron exchange events where the protons
are detected in AFP or TOTEM.}
\label{d0d0}
\end{center}
\end{figure}

It is thus possible to detect jet-gap-jet events in diffractive double pomeron
exchange processes~\cite{usb}. The idea is to tag the intact protons inside the 
AFP and CMS/TOTEM forward proton detectors~\cite{projects} located at about 
220 m from the ATLAS and CMS
interaction points on both sides. The advantage of such processes is that they
are quite clean since they are not ``polluted" by proton remnants and it is
possible to go to larger jet separation than for usual jet-gap-jet events. The
normalisation for these processes come from the fit to the D0 jet gap jet
measurements discussed in Ref.~\cite{usb}. 
The ratio between jet-gap-jet to inclusive jet events is shown
in Fig.~\ref{d0d0} requesting protons to be tagged in AFP for both samples. The ratio
shows a weak dependence as a function of jet $p_T$ (and also as a function of
the difference in rapidity between the two jets). It is worth noticing that the
ratio is about 20-30\% showing that the jet-gap-jet events are much more present
in the diffractive sample than in the inclusive one as expected.

\subsection{Exclusive diffraction}

\begin{figure}
\begin{center}
\epsfig{file=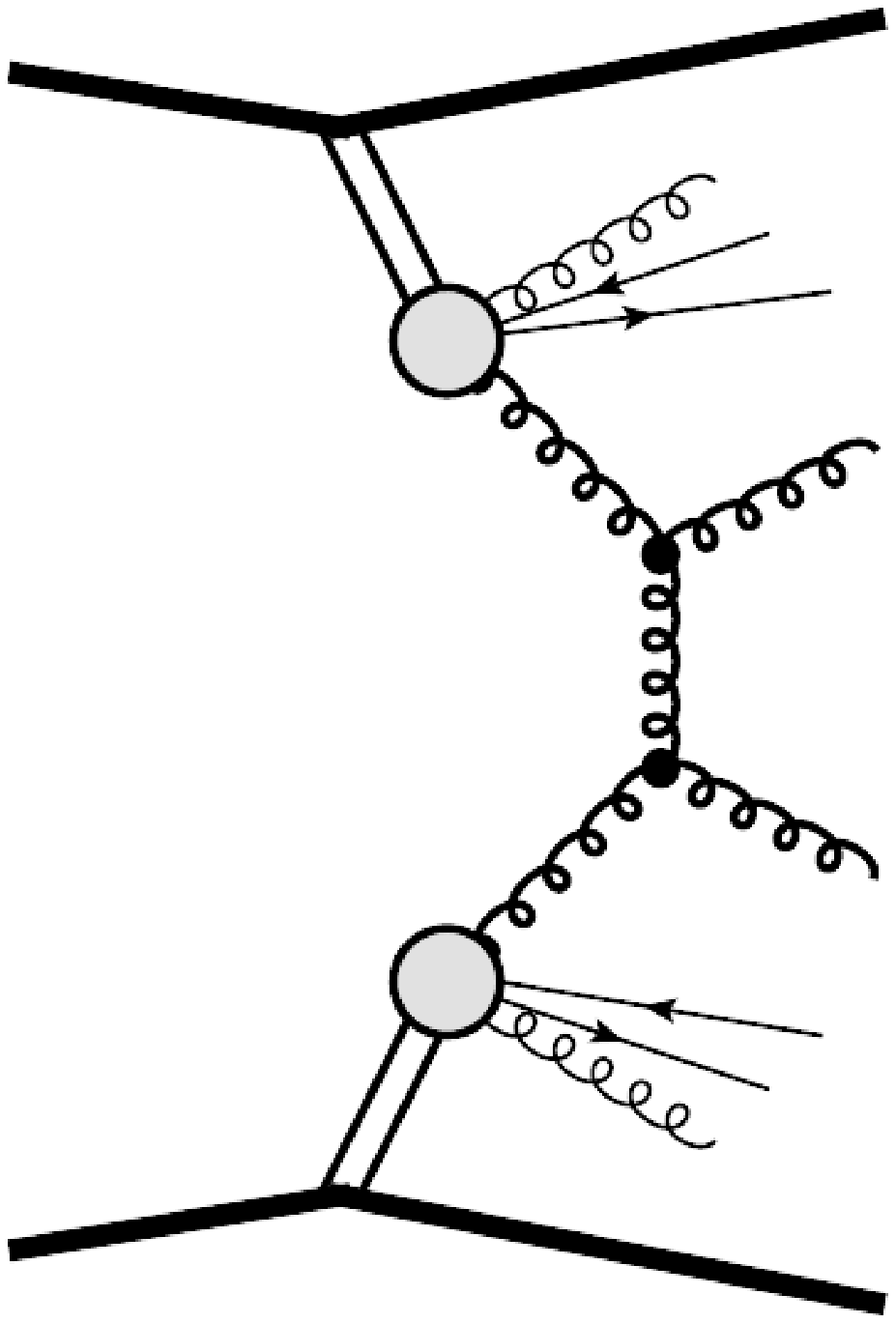,width=2.3cm}
\epsfig{file=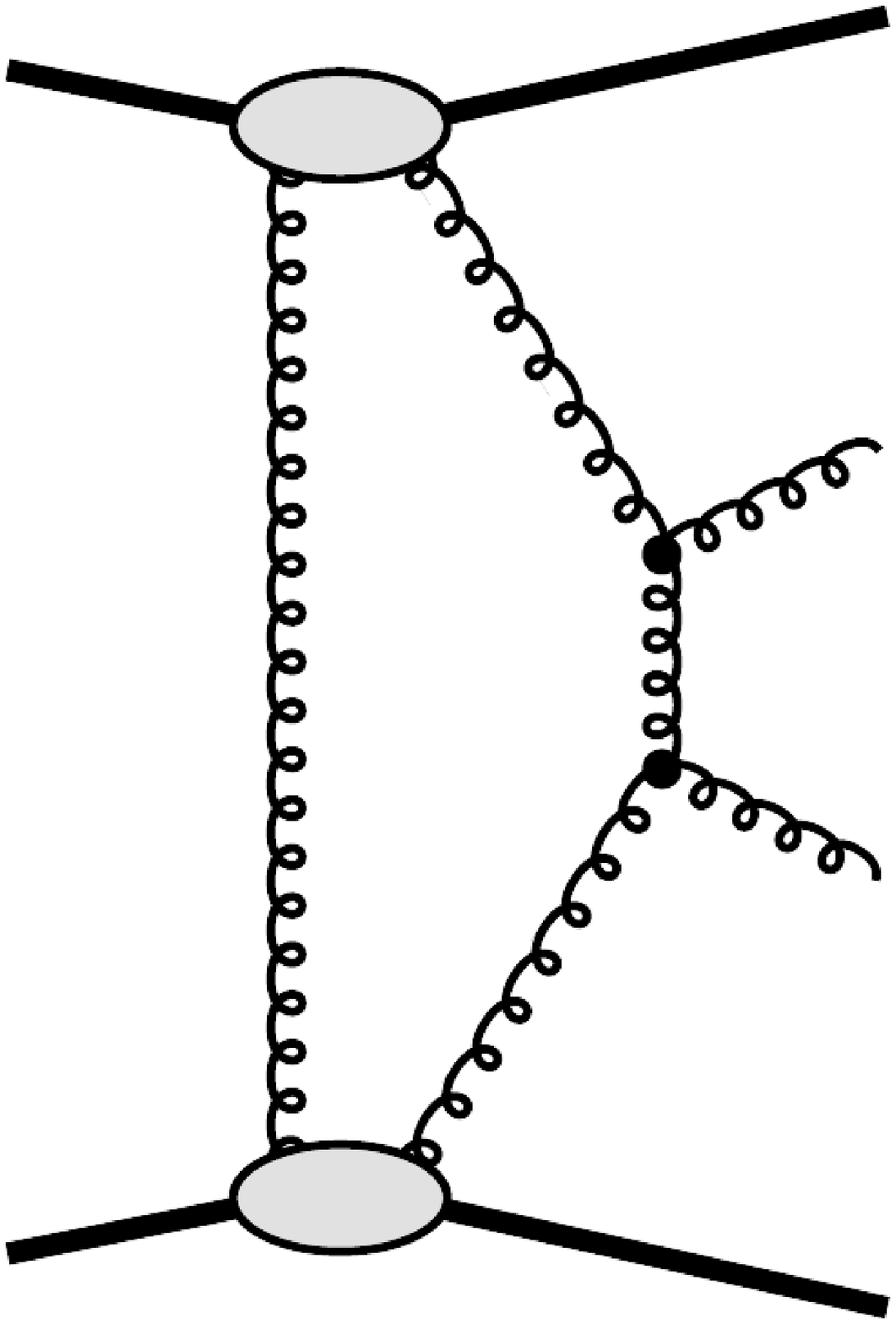,width=2.3cm}
\epsfig{file=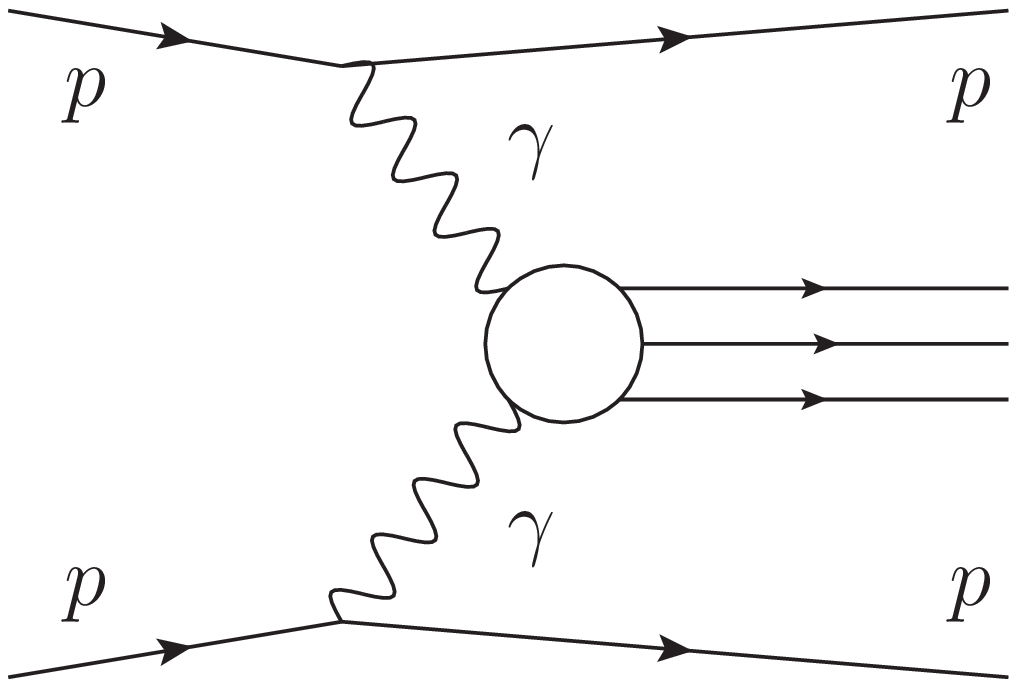,width=2.9cm}
\epsfig{file=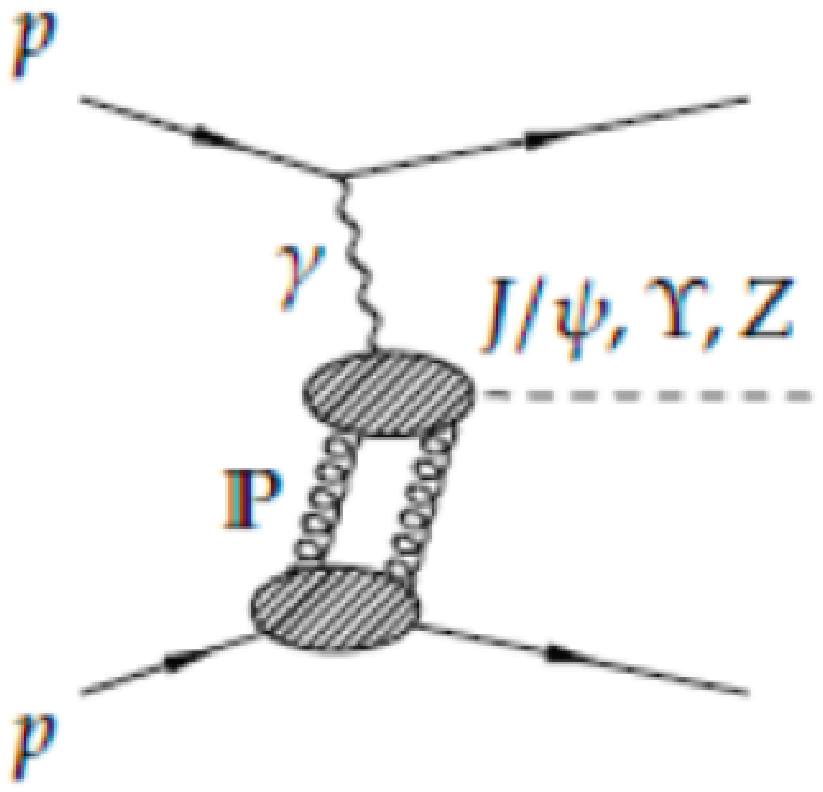,width=2.9cm}
\caption{Exclusive diffractive and photon exchange processes. The left diagram
shows the double pomeron exchange event for reference, the second one the QCD
exclusive production, the third one the production of a system $X$ via photon
exchanges, and the last one the exclusive photo-production events.}
\label{excl}
\end{center}
\end{figure}

The advantage of the exclusive diffractive and photon exchange processes
illustrated in Fig.~\ref{excl} is that all particles can be measured in the
final state. Both protons can be measured in AFP or CMS-TOTEM and the produced
particles (jets, vector mesons, $Z$ boson....) in ATLAS or CMS, and there is no
energy losses such as in the pomeron remnants as shown in Fig.~\ref{excl}, left
diagram. It is thus possible to reconstruct the properties of the object 
produced exclusively (via photon and gluon exchanges) from the tagged
proton since the system is completely constrained. It is worth mentioning that
it is also possible to constrain the background by asking the matching
between the information of the two protons and the produced object, and thus,
central exclusive production is a potential
channel for beyond standard model physics at high masses~\cite{nicolo} in the 
ATLAS and CMS collaborations as we will see in the following.

Exclusive vector mesons can be alos measured in the LHCb experiment
which recently measured for the first time the diffractive production of
charmonium~\cite{charmonium}. The Herschel scintillators are now being installed
in LHCb to enhance the coverage at high rapidities in order to get a better
control of non-exclusive background. Such channels are also sensitive to new
physics: if a medium mass resonance due to a glueball or a tetraquark state
exists, it
could lead to a bump in the invariant mass distribution of the charmonium
states.

The CMS/TOTEM experiment also performed extensive studies of possible
measurements of exclusive states at high $\beta^*$. It is worth mentioning that
the search for glueball states and the probe of the low $x$ gluon density down
to $x\sim$10$^{-4}$ will be possible. With 1 pb$^{-1}$, it will
be possible to confirm or not the existence of the unobserved possible
$f_0(1710)$ and $f_0(1500)$ decay modes and with 
5 to 10 pb$^{-1}$, the unambiguous spin determination and 
the precise measurement of cross-section times branching ratio. In addition, the
measurement of the cross section times branching ratio 
for the three $\chi_{C,0,1,2}$ states, will be performed allowing a comparison
with the
results to the LHCb measurement~\cite{lhcbchic} and the exclusive QCD
calculations~\cite{kmrphoton}

In addition, it is possible to measure the exclusive dijet production at the LHC
with about 40 fb$^{-1}$ and a pile up of 40 as was shown by the ATLAS and CT-PPS
collaborations. Despite the high level of pile up background, it is possible to
obtain a pure enough of exclusive jets that can further constrin the models of
exclusive diffractive production~\cite{projects}.


\section{Photon induced processes at the LHC and anomalous coupling studies}
In this section, we discuss some potential measurements to be performed
using proton tagging detectors at the LHC based on $\gamma$-induced processes. 
The main motivation is to explore rare events,
searching for beyond
standard model physics such as quartic anomalous couplings between photons and
$W/Z$ bosons and photons. We assume as usual in the following intact protons to be tagged 
in CMS/TOTEM or in AFP. 
These studies regained high interest recently with the observation of a potential
resonance decaying into $\gamma \gamma$ at about 750 GeV that was observed by the
ATLAS and CMS collaborations as we will see in the following~\cite{moriond,sylvain}.

In the first part of this section, we discuss the SM production of $W$ and
$\gamma$ pairs at the LHC via photon exchanges. In the second, third and fourth
sections, we discuss the sensitivities of these processes to trilinear and
quartic gauge anomalous couplings, and we finish by discussing the role of
photon-induced processes in the existence of a
potential new resonance decaying into two photons.

\subsection{Standard Model exclusive $\gamma \gamma$, $WW$ and $ZZ$ production}
\begin{figure}[t]
\hfill
\begin{minipage}[t]{.35\textwidth}

\epsfig{file=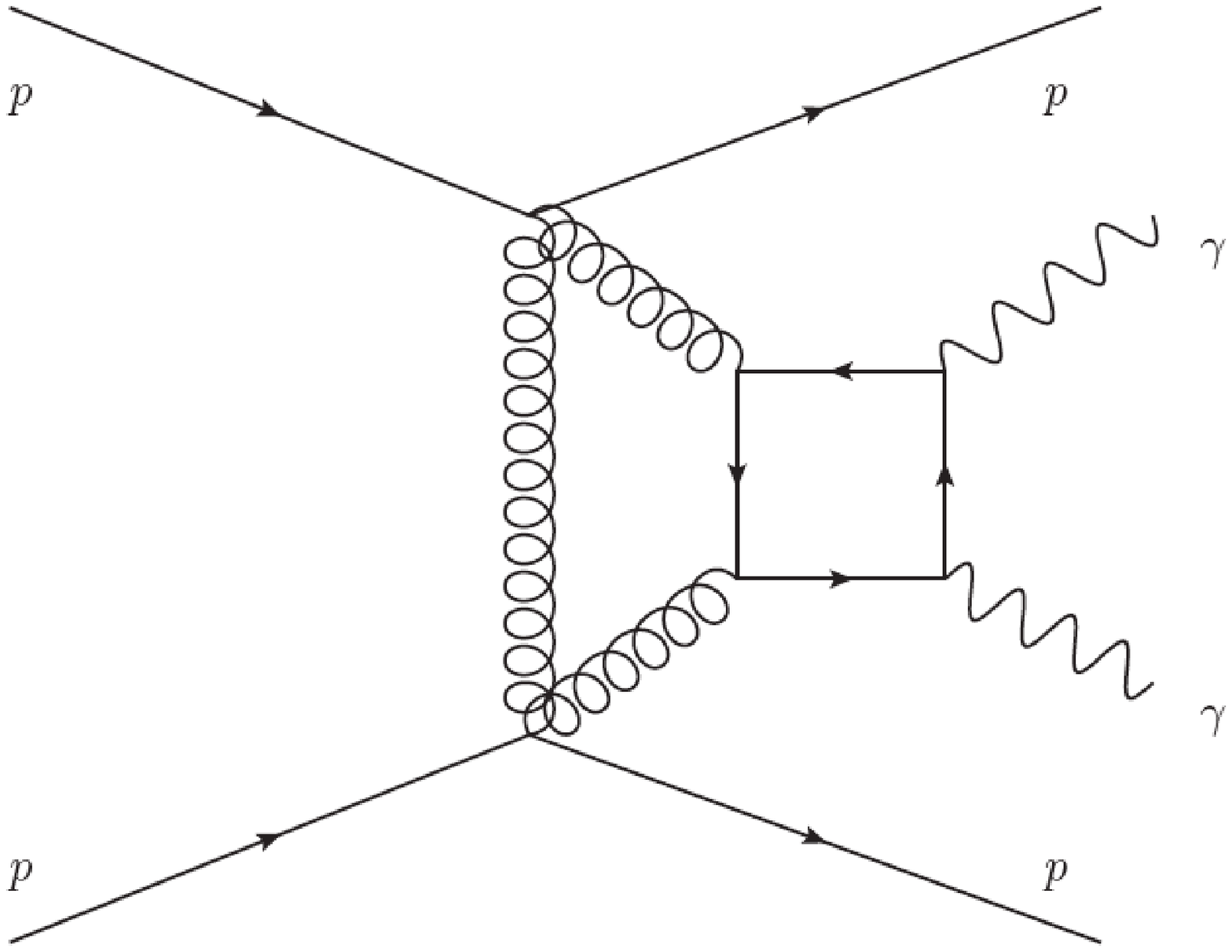,width=3.9cm} 
\caption{Diphoton QCD exclusive production. \label{fig14}}

\end{minipage}
\hfill
\begin{minipage}[t]{.45\textwidth}

\epsfig{file=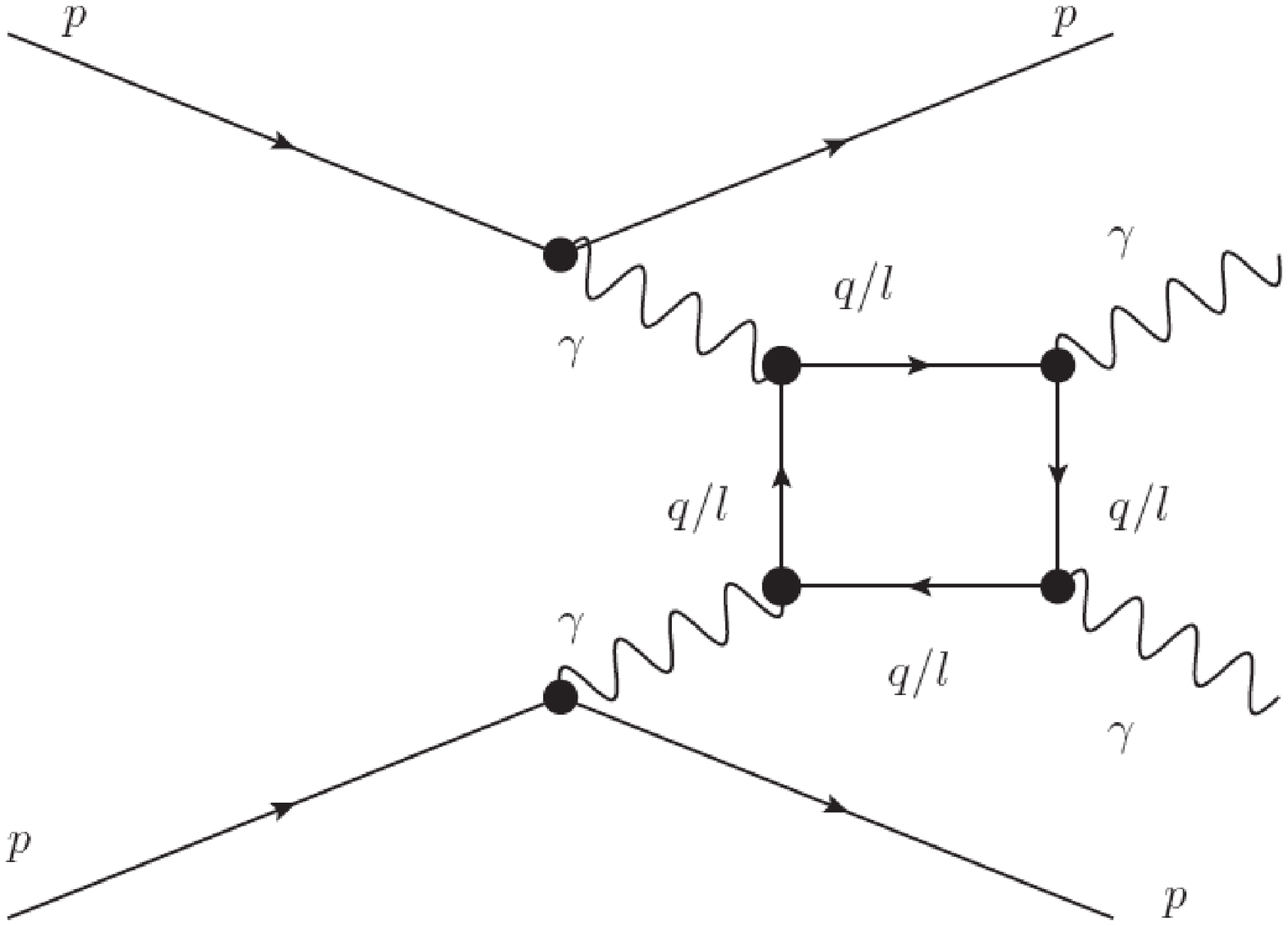,width=5.cm} 
\caption{Diphoton production via photon exchanges. \label{fig15}}

\end{minipage}
\hfill
\end{figure}

\begin{figure}
\begin{center}
\epsfig{file=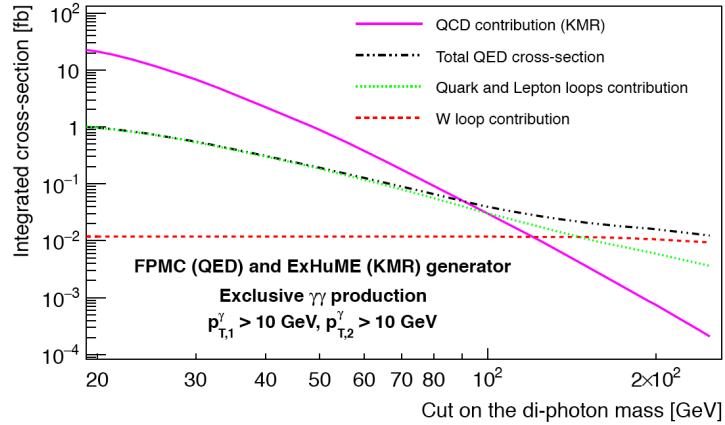,width=10.cm}
\caption{Diphoton production cross section as a function of the diphoton mass
requesting two intact protons in the final state and the photons to have 
a transverse momentum larger than 10 GeV. The QCD exclusive processes (Khoze
Martin Ryskin) in full line
dominate at low masses while QED diphoton production dominates at higher masses
(dashed lines). The QED production corresponds to diphoton production via
lepton/fermion loops (dotted line) and $W$ boson loops (dashed-dotted line).}
\label{matthias}
\end{center}
\end{figure}

In Fig.~\ref{fig14} and \ref{fig15}, we show the leading processes leading to two photons and two 
intact protons
in the final state as an example.
The first diagram (Fig.~\ref{fig14}) corresponds to exclusive QCD diphoton
production via gluon exchanges (the second gluon ensures that the exchange is
colorless leading to intact protons in the final state) and the second one 
(Fig.~\ref{fig15}) via
photon exchanges, It is worth noticing that quark, lepton and $W$ loops
need to be considered in order to get the correct SM cross section for diphoton
production as shown in Fig~\ref{matthias}. The QCD induced processes from the Khoze Martin Ryskin
model~\cite{kmrphoton} are dominant at low masses whereas
the photon induced ones (QED processes) dominate at higher diphoton 
masses~\cite{usfichet}.
It is very important to notice that the $W$ loop contribution dominates at high diphoton 
masses~\cite{usfichet} whereas this contribution is omitted in most
studies. This is the first time that we put all terms inside a MC generator,
FPMC~\cite{FPMC}.

The standard model of $W$ pair production via gluon or photon induced processes ws
also implemented in FPMC and leads to a cross section of about 96 fb at a
center-of-mass energy of 14 TeV, and it is clear that the exclusive production of $Z$
boson pairs is forbidden in the standard model.

\subsection{Triple anomalous gauge couplings}
In Ref.~\cite{usold}, we also studied the sensitivity to triple gauge
anomalous couplings at the LHC. The Lagrangian including anomalous triple gauge 
couplings $\lambda^{\gamma}$ and $\Delta\kappa^{\gamma}$ is the following
\begin{eqnarray}
   \mathcal{L} &\sim& (W^{\dagger}_{\mu\nu}W^{\mu}A^{\nu}-W_{\mu\nu}W^{\dagger\mu}A^{\nu})
   \nonumber \\
   &~&
+(1+\Delta\kappa^{\gamma})W_{\mu}^{\dagger}W_{\nu}A^{\mu\nu}+\frac{{\lambda^{\gamma}}}{M_W^2}W^{\dagger}_{\rho\mu}
   W^{\mu}_{\phantom{\mu}\nu}A^{\nu\rho} .
\end{eqnarray}
The strategy was the following: we first implemented this
lagrangian in FPMC~\cite{FPMC} and we selected the signal events when the $Z$ and $W$ bosons
decay into leptons so that the background would be negligible (the decays into
jets would be more complicated because of the high QCD dijet background). 
The difference is that the signal appears at high mass for 
$\lambda^{\gamma}$. $\Delta
\kappa^{\gamma}$ only modifies the normalization and the low mass events have 
to
be retained. The sensitivity on triple gauge anomalous couplings is a gain of
about a factor 3 with respect to the LEP limits, which represents one of the
best reaches at the LHC. 

\begin{figure}[t]
\hfill
\begin{minipage}[t]{.45\textwidth}

\epsfig{file=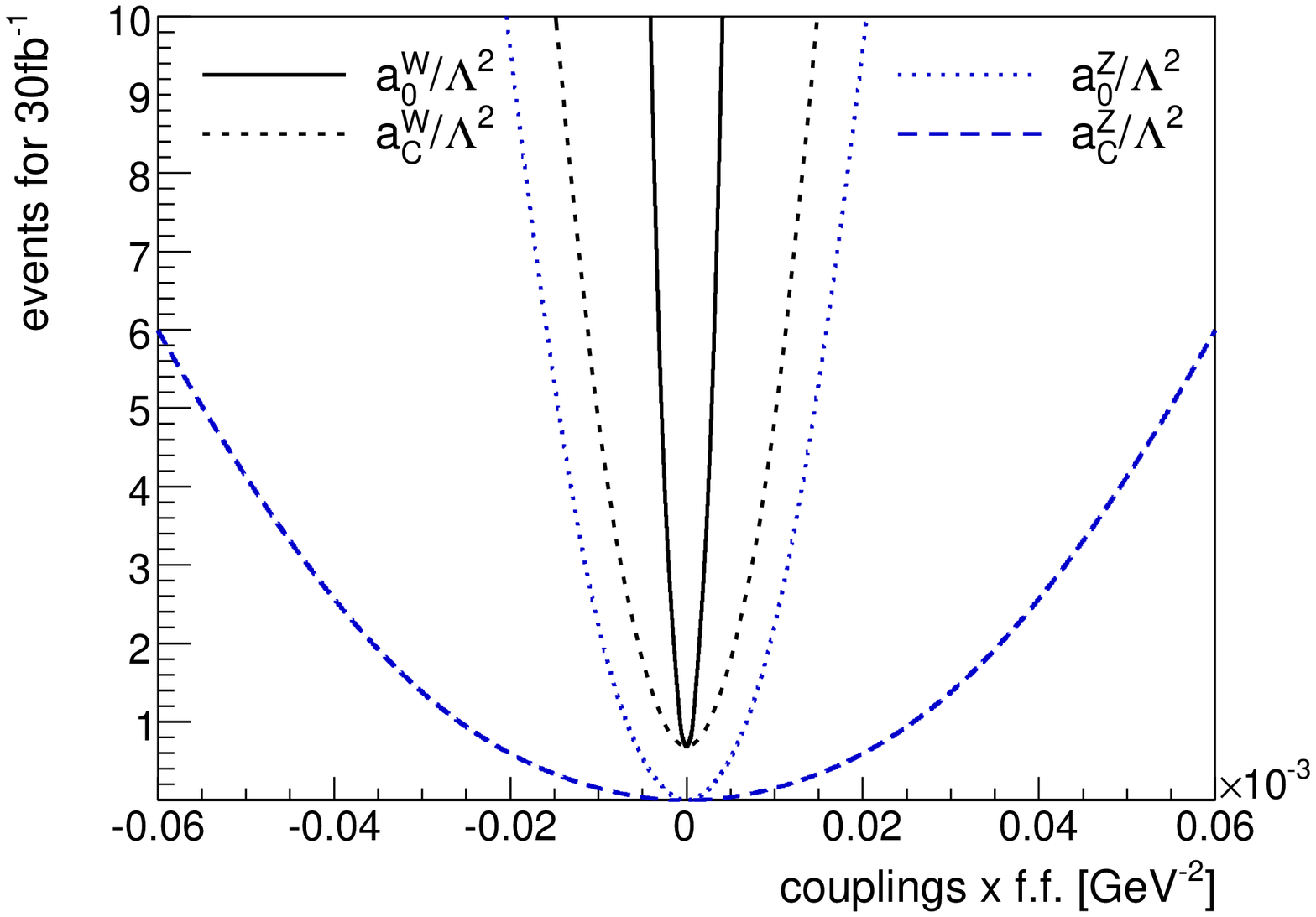,width=6.cm} 
\caption{Number of events for signal due to different values of 
anomalous couplings after all cuts (see text) for a luminosity of
30 fb$^{-1}$. \label{fig17}}

\end{minipage}
\hfill
\begin{minipage}[t]{.45\textwidth}

\epsfig{file=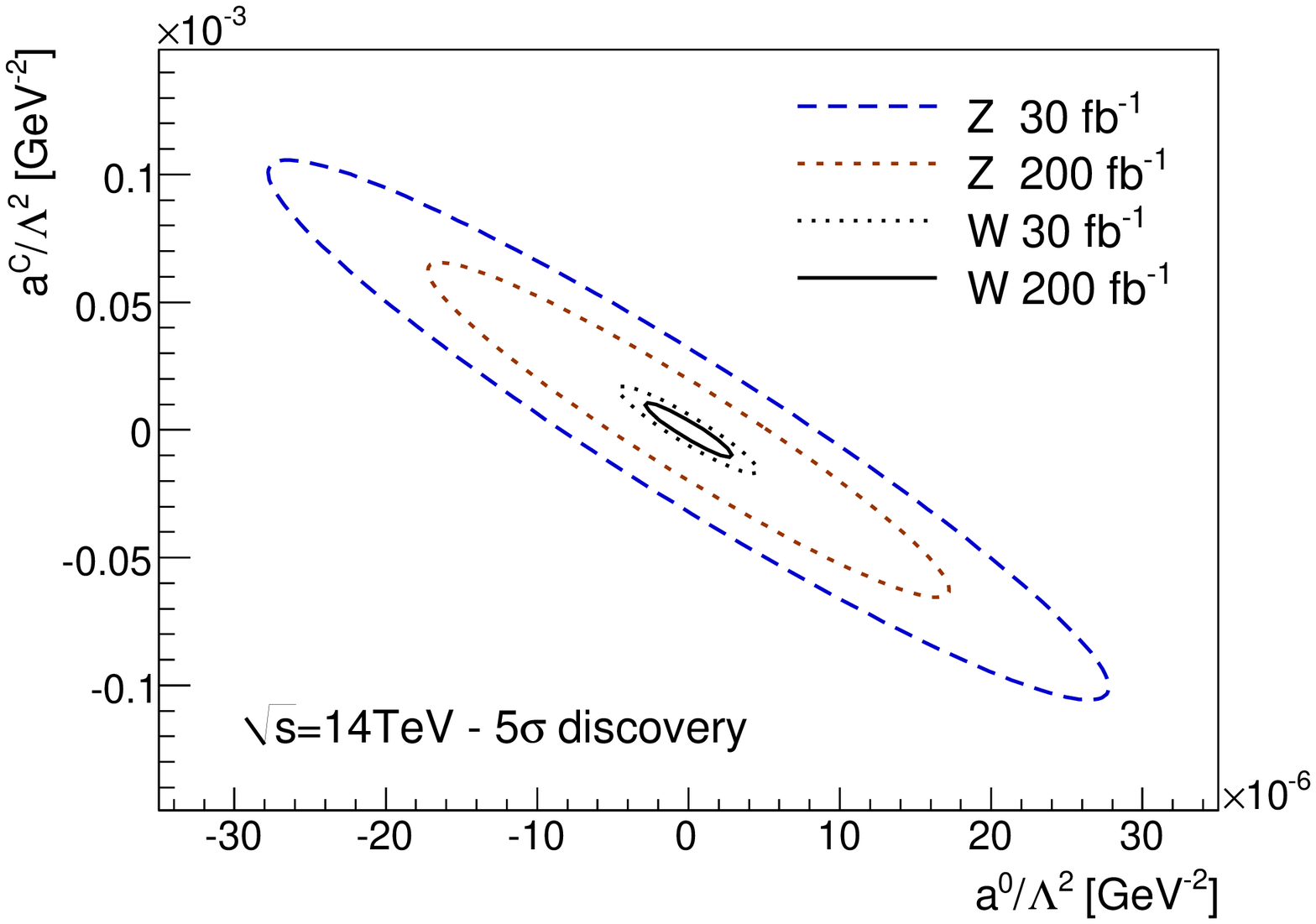,width=6.cm} 
\caption{$5\sigma$ discovery contours for all the $WW$ and $ZZ$ quartic 
couplings at $\sqrt{s}=14$ TeV for a luminosity of 30 fb$^{-1}$ and 200
fb$^{-1}$. \label{fig18}}

\end{minipage}
\hfill
\end{figure}

\subsection{Quartic $WW$ and $ZZ$ anomalous couplings}
In this section, we will study the reach at the LHC concerning the $\gamma \gamma
WW$ and $\gamma \gamma ZZ$ anomalous coupling. The principle is always similar:
we select photon-induced processes leading to two intact protons and either $W$
or $Z$ pairs in the final state. The protons are measured in CMS/TOTEM or AFP
and the $W$s and $Z$s in CMS and ATLAS.

The parameterization of the quartic couplings
based on \cite{Belanger:1992qh} is adopted. 
The cuts to select quartic anomalous gauge coupling $WW$ events are 
the following, namely $0.015<\xi<0.15$ for the
tagged protons corresponding to the AFP or CT-PPS detector at 210 and 420 m, $\met>$ 20 GeV, 
$\Delta \phi<3.13$ between the two leptons. In
addition, a cut on the $p_T$ of the leading lepton $p_T>160$ GeV and on the
diffractive mass $W>800$ GeV are requested since anomalous coupling events
appear at high mass.  
After these requirements, we expect about 0.7 background
events for an expected signal of 17 events if the anomalous coupling is about
four orders of magnitude lower than the present LEP limit~\cite{opal} ($|a_0^W / \Lambda^2| =
5.4$ 10$^{-6}$)  or two orders of magnitude lower with respect to the D0 and CMS
limits~\cite{cms} for a luminosity of 30 fb$^{-1}$. The strategy to select anomalous 
coupling $ZZ$ events is
analogous and the presence of three leptons or two like sign leptons are 
requested. 
Table 1 gives the reach on anomalous couplings at the LHC for
luminosities of 30 and 200 fb$^{-1}$ compared to the present OPAL limits from
the LEP accelerator~\cite{opal}.

Figs.~\ref{fig17} and \ref{fig18} show respectively the number of expected events for signal as a function
of the anomalous coupling value and the 5$\sigma$ discovery contours for all $WW$ and $ZZ$
anomalous couplings for 30 and 200 fb$^{-1}$.
It is possible to reach the values expected in 
extra dimension models. The tagging of the protons using the ATLAS and CMS/TOTEM
Forward
Physics detectors is likely to be the only method probe such small values of
quartic anomalous couplings.

\begin{table}
\begin{center}
   \begin{tabular}{|c||c|c|c|}
    \hline
    Couplings & 
    OPAL limits & 
    \multicolumn{2}{c|}{Sensitivity @ $\mathcal{L} = 30$ (200) fb$^{-1}$} \\
    &  \small[GeV$^{-2}$] & 5$\sigma$ & 95\% CL \\ 
    \hline
    $a_0^W/\Lambda^2$ & [-0.020, 0.020] & 5.4 10$^{-6}$ & 2.6 10$^{-6}$\\
                      &                 & (2.7 10$^{-6}$) & (1.4 10$^{-6}$)\\ \hline               
    $a_C^W/\Lambda^2$ & [-0.052, 0.037] & 2.0 10$^{-5}$ & 9.4 10$^{-6}$\\
                      &                 & (9.6 10$^{-6}$) & (5.2 10$^{-6}$)\\ \hline               
    $a_0^Z/\Lambda^2$ & [-0.007, 0.023] & 1.4 10$^{-5}$ & 6.4 10$^{-6}$\\
                      &                 & (5.5 10$^{-6}$) & (2.5 10$^{-6}$)\\ \hline               
    $a_C^Z/\Lambda^2$ & [-0.029, 0.029] & 5.2 10$^{-5}$ & 2.4 10$^{-5}$\\
                      &                 & (2.0 10$^{-5}$) & (9.2 10$^{-6}$)\\ \hline               
    \hline
   \end{tabular}
\end{center}
\caption{Reach on anomalous couplings obtained in $\gamma$ induced processes
after tagging the protons in AFP or CT-PPS compared to the present OPAL limits. The $5\sigma$ discovery and 95\%
C.L. limits are given for a luminosity of 30 and 200 fb$^{-1}$~\cite{us}} 
\end{table}

The search for quartic anomalous couplings between $\gamma$ and $W$ bosons was
performed again after a full simulation of the ATLAS and CMS detectors including pile
up~\cite{projects} assuming the protons to be tagged in AFP or CT-PPS at 210 m only. 
Integrated luminosities of 40 and 300 fb$^{-1}$ with, 
respectively, 23 or 46 average pile-up
events per beam crossing have been considered and lead to similar results. 
In order to reduce the
background, each $W$  
is assumed to decay leptonically (note that the semi-leptonic case in under study). 
The full list of background processes 
used for the ATLAS measurement of Standard Model $WW$ cross-section was
simulated, namely $t \bar{t}$, $WW$, $WZ$, $ZZ$, $W+$jets, Drell-Yan and 
single top events.

\subsection{Quartic photon anomalous couplings}

\subsubsection{Theoretical motivations}
In this section, four-photon ($4\gamma$) interactions through diphoton 
production via photon fusion with 
intact outgoing protons are considered. We will give an historical perspective
starting from a sensitivity study to anomalous $\gamma \gamma \gamma \gamma$ that
was performed before the observation of a potential resonance by the CMS and ATLAS
collaborations at a di-photon mass of 750 GeV, and we will discuss the relevance
of this observation with respect to this study in the following subsection.

In the assumption of a new physics mass scale $\Lambda$ heavier than 
the experimentally accessible 
energy $E$, all new physics manifestations can be described using 
an effective Lagrangian valid for  $\Lambda\gg E$.
Among these operators, the pure photon dimension-eight operators
\begin{eqnarray}
{\cal L}_{4\gamma}= 
\zeta_1^\gamma F_{\mu\nu}F^{\mu\nu}F_{\rho\sigma}F^{\rho\sigma}
+\zeta_2^\gamma F_{\mu\nu}F^{\nu\rho}F_{\rho\lambda}F^{\lambda\mu}
\label{zetas}
\end{eqnarray}
can induce the $\gamma \gamma \gamma \gamma$ process, highly 
suppressed in the SM~\cite{usfichet, Fichet}.
We discuss here possible new physics contributions to 
$\zeta_{1,2}^\gamma$ that can be probed and discovered at the LHC using
the forward proton detectors.   

Loops of heavy charged particles contribute to the $4\gamma$
couplings~\cite{usfichet, Fichet} as 
$\zeta_i^\gamma=\alpha^2_{\rm em} Q^4\,m^{-4}\, N\,c_{i,s}$, where
$c_{1,s}$ is related to the spin of 
the heavy particle of mass $m$ running in the loop and $Q$ 
its electric charge. 
The factor $N$  counts all additional multiplicities such as color or flavor.
These couplings scale as $\sim Q^4$ and are enhanced 
in presence of  particles with large charges. For example, certain light 
composite fermions, characteristic of composite Higgs models, have typically 
electric charges of several units. For a 500 Gev 
vector (fermion) resonance with  
$Q=3\ (4)$,
large couplings $\zeta_i^{\gamma}$ of the order of
$10^{-13}-10^{-14}$ Gev$^{-4}$ can be reached.

Beyond perturbative contributions to $\zeta_i^\gamma$ from  
charged particles,  non-renormalizable interactions of neutral particles are 
also present in common extensions of the SM.  Such theories can contain 
scalar, pseudo-scalar and spin-2 resonances that couple  to the photon
and generate  the $4\gamma$ couplings by tree-level exchange as 
$\zeta_i^\gamma=(f_{s}\, m)^{-2}\,d_{i, s}$, where
$d_{1,s}$ is related to the spin of the particle.
Strongly-coupled conformal extensions of the SM contain a scalar particle 
$(s=0^+)$, 
the dilaton. 
Even a 2 TeV dilaton can produce a sizable effective 
photon interaction, $\zeta_1^\gamma\sim 10^{-13}$\gev$^{-4}$.
These features are reproduced at large number of colors by the 
gauge-gravity correspondence in a warped extra dimension.
Warped-extra dimensions also feature Kaluza-Klein (KK) 
gravitons~\cite{Randall:1999ee}, that can induce anomalous couplings
 \cite{Fichet}
 \be
\zeta_i^\gamma= \frac{\kappa^2}{8 \tilde k^4} \,d_{i, 2}
\,,\ee 
where $\tilde k$ is the IR scale that determines the first KK graviton mass 
and $\kappa$ is a parameter that can be taken 
$\mathcal O(1)$. 
For $\kappa\sim 1$, and $m_{2}\lesssim 6$ TeV, the photon vertex can easily 
exceed $\zeta_2^\gamma\sim 10^{-14}$\gev$^{-4}$. 

\subsubsection{Experimental sensitivity to quartic four photon couplings}
The $\gamma \gamma \gamma \gamma$ process (Fig.~\ref{fig15}) can be probed 
via the detection of 
two intact protons in the forward proton detectors and two 
energetic photons in the corresponding electromagnetic 
calorimeters~\cite{usfichet}. The SM cross section of diphoton production with intact protons
is dominated by the
QED process at high diphoton mass --- and not by gluon exchanges --- and is 
thus very well known, as we saw in the previous section.

As mentioned in Ref.~\cite{ATLASECFA},
the photon identification efficiency is 
expected to be around 75\% for $p_T > 100$~GeV, with jet rejection 
factors exceeding 4000 even at high pile-up ($>$100). In addition, about 1\% of the electrons are 
mis-identified as photons. These numbers are used in the phenomenological
study presented below. For these studies, we used as an example the ATLAS
inefficiencies of identifying a photon, the mis-identification probabilities of a
jet or a lepton into a photon, and the resolution in energy and rapidities for
photons, as well as probability for conversion in lepton pairs.

\begin{figure}
\begin{center}
\includegraphics[scale=0.45]{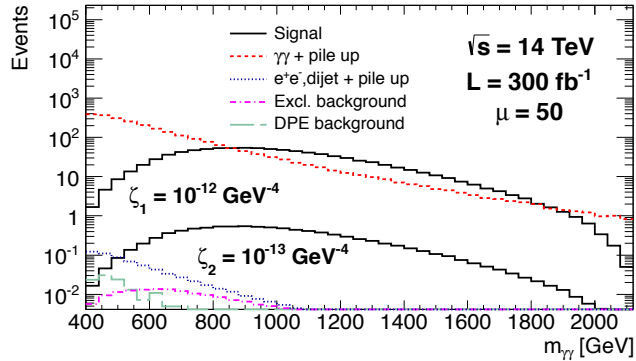}
\caption{ Diphoton invariant
mass distribution for the signal ($\zeta_{1} = 10^{-12},~10^{-13}$ Gev$^{-4}$, 
see Eq.~\ref{zetas}) and for the backgrounds 
(dominated by $\gamma\gamma$ with protons from pile-up), requesting
two protons in the forward detectors and two photons of $p_T > 50$~GeV with 
at least one converted photon in the central detector, for a luminosity of 
300 fb$^{-1}$ and an average pile-up of $\mu = 50$. \label{fig19}}
\label{fig:mass}
\end{center}
\end{figure}

\begin{table}
\small
\begin{tabular}{|c||c|c||c|c|c|c|}
\hline
Cut / Process & Signal & Signal & Excl. & DPE & DY  & $\gamma\gamma$+ \\
 & (full) & with/without & & & dijet+ & pile up \\
 & & f.f. (EFT) & & & pile up & \\
\hline
\hline
\specialcell{$[0.015<\xi_{1,2}<0.15$, \\ $p_{\mathrm{T}1,(2)}>200$
\\$(100)$ GeV]}& 65. & 18. (187.)   & 0.13 & 0.2  & 1.6           & 2968         \\
$m_{\gamma\gamma}>600$~GeV                                         & 64.   & 17.
(186.)   & 0.10 &  0    & 0.2          & 1023         \\
\specialcell{[$p_{\mathrm{T2}}/p_{\mathrm{T1}}>0.95$,\\ $|\Delta\phi|>\pi-0.01$]}    &
 64.   & 17. (186.)   & 0.10 & 0   & 0         & 80.2         \\
$\sqrt{\xi_{1}\xi_{2}s} = m_{\gamma\gamma} \pm 3\%$                & 61.   & 12.
(175.)   & 0.09 & 0   & 0         & 2.8          \\
$|y_{\gamma\gamma}-y_{pp}|<0.03$                                   & 60.   & 16.
(169.)   & 0.09 & 0   & 0         & 0            \\
\hline
\end{tabular}
\caption{Number of signal for $Q_{\rm eff}=4$, $m=340$ GeV 
and background events after 
various selections for an integrated
luminosity of 300 fb$^{-1}$ and $\mu=50$ at $\sqrt{s}=14$ TeV.
Values obtained using the corresponding EFT couplings with and without form factors are also displayed.
Excl. stands for exclusive backgrounds and DPE 
for double pomeron exchange backgrounds.
}
\label{tab:event}
\end{table}

\begin{figure*}
\includegraphics[scale=0.29]{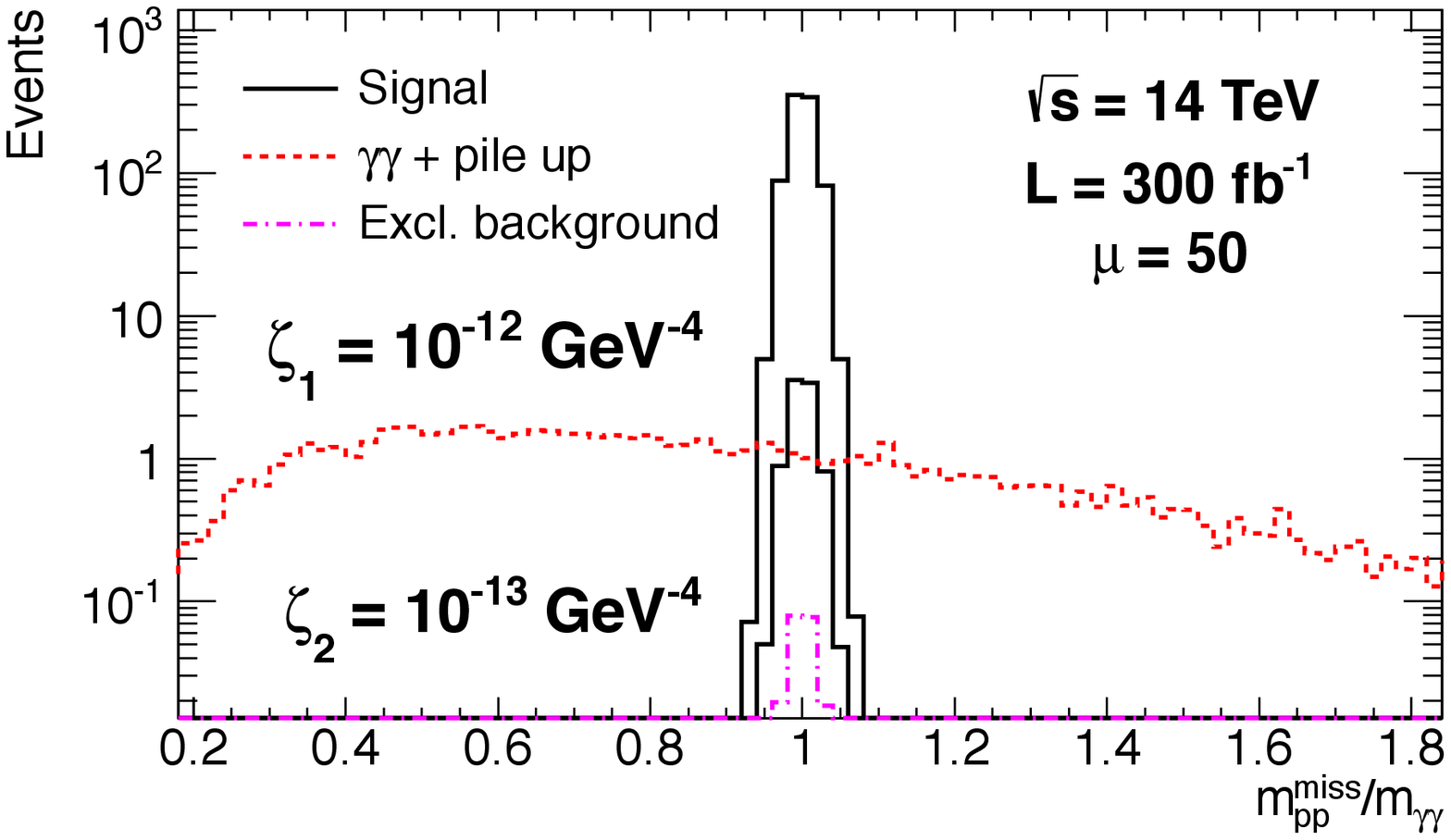}
\includegraphics[scale=0.29]{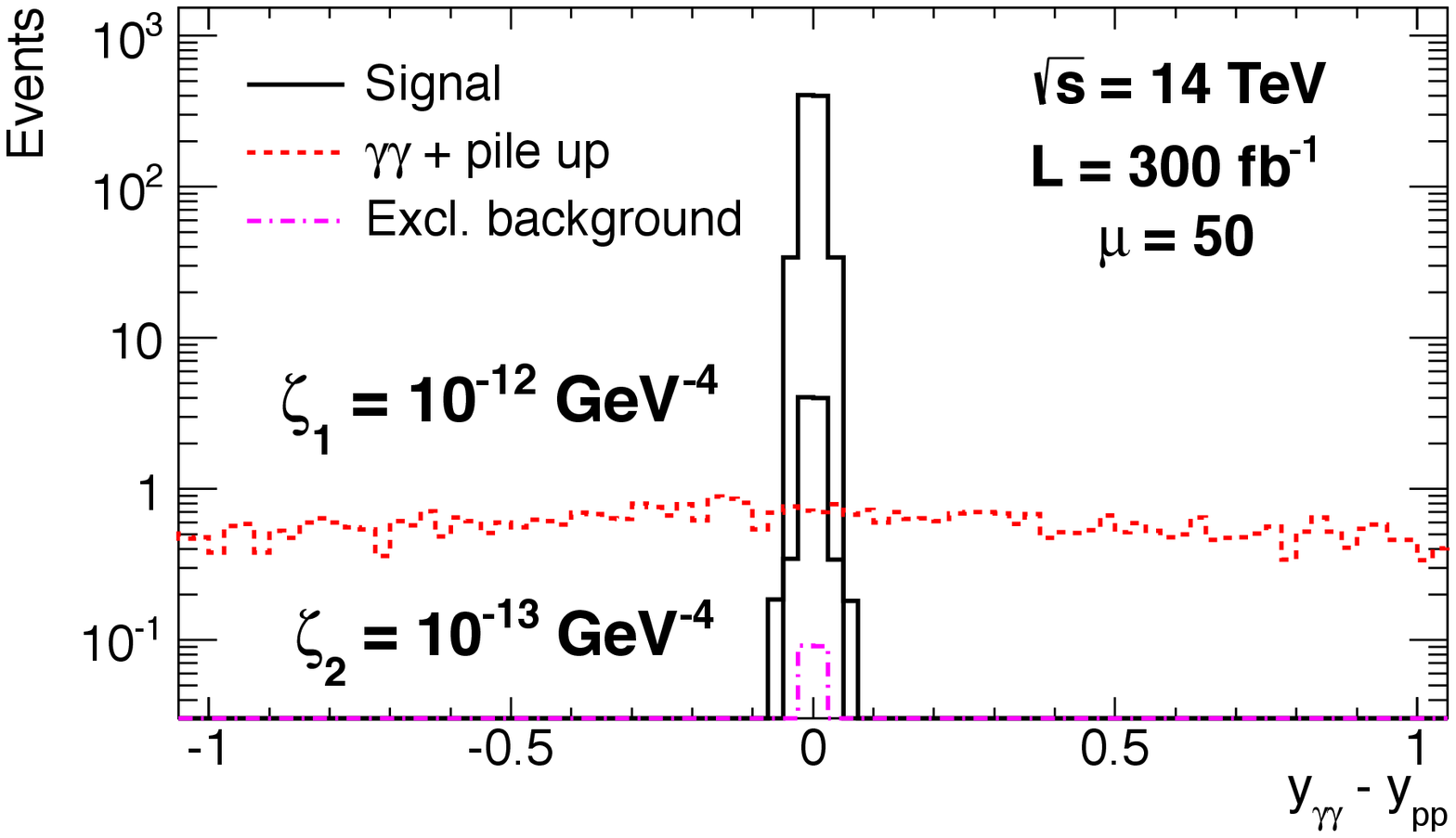}
\caption{\label{fig:massratio} Diphoton to missing proton mass ratio (left) and rapidity difference (right)
distributions for signal considering two 
different coupling values ($10^{-12}$ and $10^{-13}$\gev$^{-4}$, 
see Eq.~\ref{zetas}) and for 
backgrounds after requirements on photon $p_T$, diphoton invariant mass, $p_T$ ratio between the two photons and on the angle between the two photons. At least one converted photon is required. The integrated luminosity 
is 300~fb$^{-1}$ and the average pile-up is $\mu=50$. \label{fig20} }
\end{figure*}

As for the previous studies, the anomalous $\gamma \gamma \gamma \gamma$ 
process has been 
implemented in the Forward Physics Monte Carlo (FPMC) generator~\cite{FPMC}.
The FPMC generator was also used to simulate the background 
processes giving rise to two intact protons accompanied by two photons, 
electrons or jets that can mimic the photon signal. Those include exclusive 
SM production of $\gamma \gamma \gamma \gamma$ via lepton and quark boxes and 
$\gamma\gamma\rightarrow e^{+}e^{-}$. The central exclusive 
production of $\gamma\gamma$ via two-gluon exchange, not present in FPMC,
was simulated using ExHuME~\cite{Monk:2005ji}. This series of backgrounds is 
called
``Exclusive" in Table~\ref{tab:event} and Figs.~\ref{fig19}, \ref{fig20}.
FPMC was also used to produce $\gamma\gamma$, 
Higgs to $\gamma\gamma$ and dijet productions via double pomeron exchange 
(called
DPE background in Table~\ref{tab:event} and Fig.~\ref{fig19}).
Such backgrounds tend to be softer than the signal and can be suppressed with 
requirements on the transverse momenta of the photons and the diphoton invariant 
mass. 
In addition, the final-state photons of the signal are typically back-to-back 
and have 
about the same transverse momenta. Requiring a large azimuthal 
angle $|\Delta \phi| > \pi -0.01$ between the two photons and a 
ratio $p_{T,2} / p_{T,1} > 0.95$ greatly reduces the contribution of 
non-exclusive processes.

Additional background processes include the quark and gluon-initiated 
production of two photons, two jets and Drell-Yan processes leading to two 
electrons. The two intact 
protons arise from pile-up interactions (these backgrounds are called 
$\gamma\gamma$ + pile-up and e$^{+}$e$^{-}$, dijet + pile-up in 
Table~\ref{tab:event}).These events were produced using HERWIG~\cite{herwig} and
PYTHIA~\cite{pythia}. 
The pile-up background is further suppressed by requiring the proton 
missing invariant mass to match the diphoton invariant mass within 
the expected 
resolution and the diphoton
system rapidity and the rapidity of the two protons to be similar
(see Fig. 17).

The number of expected signal and background events passing respective 
selections is shown in 
Table~\ref{tab:event} for an integrated luminosity of 300 fb$^{-1}$\ 
for a center-of-mass energy of 14\,TeV~\cite{usfichet}.
Exploiting the full event kinematics with the forward proton detectors 
allows to completely suppress the background with a signal selection 
efficiency after the acceptance cuts exceeding 70\%. Tagging the protons
is absolutely needed to suppress the $\gamma \gamma$ + pile-up events.
Further background reduction is even possible by requiring the photons 
and the protons to originate from the same vertex that provides an additional 
rejection 
factor of 40 for 50 pile-up interactions, showing the large margin on the 
background suppression.
A similar study 
at a higher pile-up of 200 was performed 
and led to a very small background.
The sensitivities
on photon quartic anomalous couplings are given in Table~\ref{sensitivities}.
The sensitivity extends up to $7\cdot10^{-15}$ GeV$^{-4}$ allowing us to probe further the models of new
physics described above.

\begin{table}
\begin{center}
\begin{tabular}{|c||c|c||c|c||c|}
\hline
Luminosity & 300~\fbi & 300~\fbi & 300~\fbi & 300~\fbi & 3000~\fbi \\
\hline
 pile up ($\mu$) & 50 & 50 & 50 & 50 & 200 \\
\hline
\hline
coupling & $\ge$~1 conv. $\gamma$ & $\ge$~1 conv. $\gamma$ & all $\gamma$ & all $\gamma$  & all $\gamma$\\
(GeV$^{-4}$) & 5 $\sigma$ & 95\% CL & 5 $\sigma$ & 95\% CL & 95\% CL \\

\hline
$\zeta_1$~f.f.   &  $8\cdot10^{-14}$   & $5\cdot10^{-14}$   & $4.5\cdot 10^{-14}$ & $3\cdot 10^{-14}$ & $2.5\cdot10^{-14}$ \\
$\zeta_1$~no f.f.&  $2.5\cdot10^{-14}$ & $1.5\cdot10^{-14}$ & $1.5\cdot10^{-14}$ & $9\cdot10^{-15}$  & $7\cdot10^{-15}$\\
\hline
$\zeta_2$~f.f.   &  $2\cdot10^{-13}$   & $1\cdot10^{-13}$   & $9\cdot10^{-14}$ & $6\cdot10^{-14}$  & $4.5\cdot10^{-14}$ \\
$\zeta_2$~no f.f.&  $5\cdot10^{-14}$   & $4\cdot10^{-14}$   & $3\cdot10^{-14}$& $2\cdot10^{-14}$  & $1.5\cdot10^{-14}$ \\
\hline

\end{tabular}
\end{center}

\caption{5\,$\sigma$ discovery and 95\% CL exclusion limits on $\zeta_1$ and $\zeta_2$ 
couplings in\gev$^{-4}$ (see Eq.~\ref{zetas}) with
and without form factor (f.f.), requesting at least one converted photon ($\ge$~1 conv. $\gamma$) or
not (all $\gamma$). All sensitivities are given for 300 fb$^{-1}$
and $\mu=50$  pile up events (medium luminosity LHC) except for the numbers of the last column which are given for 3000
fb$^{-1}$ and $\mu=200$  pile up events (high luminosity LHC). }
\label{sensitivities}
\end{table}

We also performed a full amplitude calculation in Ref.~\cite{usfichet} that avoids the
dependence on the choice of form factors needed in order to avoid quadratic
divergences of scattering amplitudes. Sensitivities were found to be similar
leading to possible discoveries of vector or fermions at high masses and high
effective charges.

If discovered at the LHC, $\gamma \gamma \gamma
\gamma$ quartic anomalous couplings would be a major discovery related
to the existence of extra dimensions in the universe as an example. 
In addition, it might be inveestigated if there could be a link with
some experiments in atomic physics, for instance the intrication experiments that  
might be interpreted via the existence of 
extra dimensions.
Further more, it is clear that extra dimensions might be relevant also for the
fast expansion of the universe within inflation models.

\subsection{Photon quartic anomalous couplings and the potential presence of a
resonance in the di-photon mass spectrum at 750 GeV observed by the ATLAS and CMS
collaborations}

Recently, the CMS and ATLAS collaborations anounced the presence of a small excess in
the di-photon mass spectrum for a mass of about 750 GeV at a center-of-mass energy
of 13 TeV. While it is still too early
to know if this excess is real or a statistical fluctuation, it is important to know
how proton forward detectors might be able to give information about the production
mechanism. Two experimental facts were observed: the excess is not present or very
small at a center-of-mass energy of 8 TeV, and not seen in the di-jet channel at 13
TeV. If processes were gluon-induced, we would expect the di-jet cross section to be
of the order of 1 pb since the ratio of the di-jet to di-photon cross section is of
the order of $\Gamma_{gg}/\Gamma_{\gamma \gamma} = \alpha_S^2 / \alpha^2 \sim 200$.
Since no excess is observed in the di-jet channel, it seems natural to consider
photon-induced processes (it is clear that the potential resonance might also be
produced in a combination of photon and gluon-induced processes). 

In Section 3.1, we already showed that the di-photon production via photon exchanges
completely dominates the one via gluon exchanges at high di-photon masses. This means
that we are sure that di-photon production is photon-induced is we tag the intact
protons in the final state for a di-photon resonance of about 750 GeV.

Concerning the excess that has been presently observed, the forward protons detectors
were not included in data taking. This means that for most of the events, protons were
destroyed in the final state corresponding to inelastic events. The cross section
of the $pp \rightarrow R \rightarrow \gamma \gamma X$ reads~\cite{sylvain}
\begin{eqnarray}
\sigma_{pp\rightarrow \gamma\gamma X}=  ( 7.3~{\rm fb})\, 
\left(\frac{5\, {\rm TeV}}{f_{\gamma}}\right)^4 
\left(\frac{45~{\rm GeV}}{\Gamma_{\rm tot}}\right) 
\left(\frac{r_{\rm inel}}{20}\right) 
\end{eqnarray}
where $f_{\gamma}$, $\Gamma_{\rm tot}$, $r_{inel}$ are respectively the $\gamma \gamma
R$ coupling, the width of the resonance, and the ration between the inelastic and
elastic contributions. As determined by data, $f_{\gamma}$ is of the order of 5 TeV,
and the width $\Gamma_{tot}$ of the order of 45 GeV. The $r_{inel}$ parameters is
about 20 with a large uncertainty. More recent stduies lead to a better
determination of this factor~\cite{lucian}.
The fact that the resonance can be produced at 13
TeV but almost not at 8 TeV is related to the probability to emit quasi-real photons from
the proton that can couple to the resonance $R$. The phase space producing a 750 GeV
resonance at a center-of-mass energy of 8 TeV is much reduced compared to 13 TeV. We
estimated this factor to be between 2.4 and 3.9. 

Observing di-photon exclusive production and tagging the intact protons in the final
state will allow being certain that these processes are photon-induced. As we mention
already, this is a background-free experiment, which means that the observation of 5
events is enough to obtain a 5$\sigma$ discovery. We predict the following cross
section for the $pp \rightarrow p \gamma \gamma p$ process
\begin{eqnarray}
\sigma_{pp\rightarrow \gamma\gamma pp}=  ( 0.23~{\rm fb})\, 
\left(\frac{5\, {\rm TeV}}{f_{\gamma}}\right)^4 
\left(\frac{45~{\rm GeV}}{\Gamma_{\rm tot}}\right) r_{fs} 
\end{eqnarray}
where $f_{fs}$ is the survival probability that can be estimated to be of the order of
0.8. About 20 fb$^{-1}$ is thus necessary to obtain a 5$\sigma$ discovery in this
channel.

In addition to the $\gamma \gamma$ channel, we predict a possible significant
production of $ZZ$, $WW$, and $Z \gamma$ events and it would be also interesting to
look in these channels with tagged protons as well.

\section{Conclusion}
In this report, we detailled the interest of tagging the intact protons to
study in detail the pomeron structure in terms of quarks and gluons
and $WW$, $ZZ$ and $\gamma \gamma$ productions via photon exchanges.
Unprecedented sensitivities can be achieved at the LHC in the CMS-TOTEM and ATLAS
experiments on quartic anomalous couplings, especially on $\gamma \gamma \gamma
\gamma$ couplings, that will lead to one of the best sensitivity on
extra dimensions at the LHC, and to potential discoveries if the existence of the
di-photon resonance is confirmed.



\begin{thebibliography}{99}

\bibitem{dipole} H. Navelet, R. peschanski, C. Royon, Phys.Lett. B{\bf 366} 
(1996) 329; H. Navelet, Robert B. Peschanski, C. Royon, S. Wallon,
Phys.Lett. B{\bf 385} (1996) 357; 	
A. Bialas, R. Peschanski, C. Royon, Phys.Rev. D{\bf 57} (1998) 6899.  

\bibitem{dglap} 
G.Altarelli and G.Parisi,
{\it Nucl. Phys.} {\bf B126}  18C (1977) 298;
V.N.Gribov and L.N.Lipatov, {\it Sov. Journ. Nucl. Phys.} (1972) 438 and 675;
Yu.L.Dokshitzer, {\it Sov. Phys. JETP.} {\bf 46} (1977) 641.


\bibitem{fith1} H1 Collaboration, arXiv:hep-ex/0606004; preprint hep-ex/0606003;
ZEUS Collaboration,
Nucl. Phys. {\bf  B 713} (2005) 3;
C. Royon, L. Schoeffel, J. Bartels, H. Jung, R. Peschanski,
Phys.Rev. D{ \bf 63} (2001) 074004;
C. Royon, L. Schoeffel, S. Sapeta, R. Peschanski, E. Sauvan,
Nucl.Phys. B {\bf 781} (2007) 1.

\bibitem{bfkl} V. S. Fadin, E. A. Kuraev, L. N. Lipatov, Phys. Lett. B{\bf 60} (1975) 50; I. I. Balitsky, L. N. 
Lipatov, Sov.J.Nucl.Phys. {\bf 28} (1978) 822.


\bibitem{yellow} Report from the LHC Forward Physics Working Group, N. Cartiglia
and C. Royon Editors, to be published in J. Phys. G.

\bibitem{lhcf} LHCf Coll., for results see hep.fi.infn.it/LHCf/.

\bibitem{lhcb} LHCb Coll., for results see
http://lhcb.web.cern.ch/lhcb/Physics-Results/LHCb-Physics-Results.

\bibitem{alice} ALICE Coll., for results see 
https://twiki.cern.ch/twiki/bin/view/ALICEpublic /ALICEPublicResults


 
\bibitem{qcd} C. Marquet, C. Royon, M. Saimpert, D. Werder, Phys. Rev. D{\bf 88}
(2013) 7, 074029.



\bibitem{nicolo}  
A special issue on central exclusive production in hadron-hadron collisions, 
Int.J.Mod.Phys. A{\bf 29}, number 28 (2014), Editors: M.
Albrow, V. Khoze, C. Royon.

\bibitem{projects} ATLAS Coll., CERN-LHCC-2011-012; TOTEM Coll.,
CERN-LHCC-2014-020; 
CMS and TOTEM Coll., CERN-LHCC-2014-021.


\bibitem{reviewdif}  
A special issue on Diffractive Processes in Lepton-Hadron and Hadron-Hadron
Collisions, Int. J. Mod. Phys. A {\bf 30}, number 08 (2015),
 Editor: C. Royon.


\bibitem{matthias} C. Royon, M. Saimpert, O. Kepka, R. Zlebcik,
Acta Physica Polonica B, Proceedings supplement, Vol. 7, Number 4 (2014) 735.

\bibitem{cdfus}O. Kepka, C, Royon, Phys.Rev. D{\bf 76} (2007) 034012.

\bibitem{annabelle} A. Chuinard, C. Royon, R. Staszewski, JHEP 1604 (2016) 092. 

\bibitem{usb} C. Marquet, C. Royon, M. Trzebinski,  R. Zlebcik,
Phys.Rev. D{\bf 87} (2013) no.3, 034010; O. Kepka, C. Marquet, C. Royon,
Phys.Rev. D{\bf 83} (2011) 034036.

\bibitem{herwig} G. Corcella et al., arXiv:hep-ph/0210213.


\bibitem{charmonium} LHCb Coll., preprint  arXiv:1407.5973.

\bibitem{lhcbchic}  LHCb Coll., see
https://twiki.cern.ch/twiki/bin/view/LHCb/PrelimExclDiMu

\bibitem{kmrphoton}
  V.~A.~Khoze, A.~D.~Martin and M.~G.~Ryskin,
  Eur.\ Phys.\ J.\  C {\bf 23} (2002) 311.
  
\bibitem{moriond} CMS Coll., CMS-PAS-EXO-15-004; ATLAS Coll.,
ATLAS-CONF-2015-081.


\bibitem{usfichet} S. Fichet, G. von Gersdorff, O. Kepka, B. Lenzi, C. Royon, M.
Saimpert, Phys. Rev. D{\bf 89} (2014) 114004;
S. Fichet, G. von Gersdorff, B. Lenzi, C. Royon, M. Saimpert,.
JHEP {\bf 1502} (2015) 165.


\bibitem{sylvain} S. Fichet, G. von Gersdorff, C. Royon, Phys.Rev. 
D{\bf 93} (2016) no.7, 075031; S. Fichet, G. von Gersdorff, C. Royon, 
arXiv:1601.01712, Phys.Rev.Lett. {\bf 116} (2016) no.23, 231801.


\bibitem{FPMC}  
M. Boonekamp, A. Dechambre, V. Juranek, O. Kepka, M. Rangel, 
C. Royon, R. Staszewski, e-Print: arXiv:1102.2531;    
M. Boonekamp, V. Juranek, O. Kepka, C. Royon ``Forward Physics Monte Carlo'',
      ``Proceedings of the workshop: HERA and the LHC workshop series on the
      implications of HERA for LHC physics,'' arXiv:0903.3861 [hep-ph].



\bibitem{usold} 
O.~Kepka and C.~Royon,
Phys.\ Rev.\  D {\bf 78} (2008) 073005.




\bibitem{Belanger:1992qh}
  G.~Belanger and F.~Boudjema,
  Phys.\ Lett.\  B {\bf 288} (1992) 201.
 
\bibitem{cms}  CMS Coll., JHEP {\bf 07} (2013) 116; D0. Coll., Phys. Rev. D {\bf
88} (2013) 012005. 


\bibitem{opal}
  G.~Abbiendi {\it et al.}  [OPAL Collaboration],
  Phys.\ Rev.\  D {\bf 70} (2004) 032005
  [arXiv:hep-ex/0402021].
 

\bibitem{Fichet}
  S.~Fichet and G.~von Gersdorff,
  preprint arXiv:1311.6815.;
  R.~S.~Gupta,
  Phys.\ Rev.\ D {\bf 85} (2012) 014006.

\bibitem{Randall:1999ee} 
  L.~Randall and R.~Sundrum,
  Phys.\ Rev.\ Lett.\  {\bf 83} (1999) 3370.
 

\bibitem{ATLASECFA} ATLAS Coll., ATL-PHYS-PUB-2013-009;
ATLAS Coll., JINST, Vol. 3 (2008) S08003.


\bibitem{Monk:2005ji}
  J.~Monk and A.~Pilkington,
  Comput.\ Phys.\ Commun.\  {\bf 175} (2006) 232; V.A. Khoze, A.D. Martin, M.G. Ryskin,
  Eur.Phys.J. C {\bf 55} (2008) 363. 



\bibitem{pythia}
  T.~Sjostrand, S.~Mrenna and P.~Z.~Skands,
  Comput.\ Phys.\ Commun.\  {\bf 178} (2008) 852.


\bibitem{us} E. Chapon, O. Kepka, C. Royon, Phys. Rev. {\bf D81} (2010) 074003;
J. de. Favereau et al., preprint arXiv:0908.2020; O.~Kepka and C.~Royon,
Phys.\ Rev.\  D {\bf 78} (2008) 073005.


\bibitem{lucian} 	
L.A. Harland-Lang, V.A. Khoze, M.G. Ryskin,
JHEP 1603 (2016) 182; A.Martin, M. G. Ryskin, J. Phys. G{\bf 43}
(2016) no 4, 04LT02.

\end{thebibliography}
\end{document}